\documentclass[longauth]{aa}

\usepackage[utf8]{inputenc}
\usepackage[T1]{fontenc}
\usepackage[english]{babel}

\DeclareRobustCommand{\VAN}[3]{#2}
\let\VANthebibliography\thebibliography
\def\thebibliography{\DeclareRobustCommand{\VAN}[3]{##3}\VANthebibliography}

\usepackage{graphicx}	
\graphicspath{{./figures/}}
\usepackage{amsmath}	

\usepackage{color}
\usepackage{booktabs}
\usepackage{tabularx}
\usepackage{multicol}
\usepackage{multirow}
\usepackage{hyperref}

\makeatletter
\renewcommand*\aa@pageof{, page \thepage{} of \pageref*{LastPage}}
\makeatother

\def\aap{A\&A}

\def\apjs{ApJS}
\def\apjl{ApJL}

\newcommand{\lyb}{Ly$\beta$}
\newcommand{\lya}{Ly\texorpdfstring{$\alpha$}{alpha}}

\newcommand{\squeze}{\texttt{SQUEzE}}
\newcommand{\siiv}{\ion{Si}{IV}}
\newcommand{\civ}{\ion{C}{IV}}

\newcommand{\ciii}{\ion{C}{III]}}

\newcommand{\mgii}{\ion{Mg}{II}}
\newcommand{\oii}{\ion{O}{[II]}}

\newcommand{\hb}{H$\beta$}
\newcommand{\oiii}{\ion{O}{[III]}}
\newcommand{\ha}{H$\alpha$}
\newcommand{\angs}{\mbox{\normalfont\AA}}

\begin{document}


\title{The miniJPAS survey quasar selection IV: Classification and redshift estimation with SQUEzE}
\titlerunning{The miniJPAS survey quasar selection IV}

\author{Ignasi P\'erez-R\`afols
    \inst{1}$^{,}$\inst{2}$^{,}$\inst{3}\fnmsep\thanks{Email: iprafols@fqa.ub.edu}
    \and
    L. Raul Abramo\inst{4}
    \and
    Gin\'es Mart\'inez-Solaeche\inst{5}
    \and
    Matthew M. Pieri\inst{6}
    \and
    Carolina Queiroz\inst{4}
    \and
    Nat\'alia V.N. Rodrigues\inst{4}
    \and
    Silvia Bonoli\inst{7}$^{,}$\inst{8}
    \and 
    Jon\'as Chaves-Montero\inst{2}$^{,}$\inst{7}
    \and
    Sean S. Morrison\inst{6}$^{,}$\inst{9}
    \and
    Jailson Alcaniz\inst{10}
    \and
    Narciso Benitez\inst{11}
    \and
    Saulo Carneiro\inst{12}
    \and 
    Javier Cenarro\inst{13}
    \and
    David Cristóbal-Hornillos\inst{14}
    \and
    Renato Dupke\inst{10}
    \and
    Alessandro Ederoclite\inst{13}
    \and
    Rosa M. González Delgado\inst{11}
    \and
    Antonio Hernán-Caballero\inst{13}
    \and
    Carlos López-Sanjuan\inst{13}
    \and 
    Antonio Marín-Franch\inst{13}
    \and
    Valerio Marra\inst{15}$^{,}$\inst{16}$^{,}$\inst{17}
    \and
    Claudia Mendes de Oliveira\inst{18}
    \and
    Mariano Moles\inst{14}
    \and
    Laerte Sodré Jr.\inst{18}
    \and
    Keith Taylor\inst{19}
    \and
    Jesús Varela\inst{14}
    \and
    Héctor Vázquez Ramió\inst{13}
    }

\authorrunning{I. Pérez-Ràfols et al.}

   \institute{
    Departament de Física Quàntica i Astrofísica. Institut de Ci\`encies del Cosmos, Universitat de Barcelona (UB-IEEC), Mart\'i i Franqu\`es 1, E-08028 Barcelona, Catalonia, Spain. \email{iprafols@fqa.ub.edu}
    \and
    Institut de Física d’Altes Energies (IFAE), The Barcelona Institute of Science and Technology, 08193 Bellaterra (Barcelona), Spain
    \and
    Institut d'Estudis Espacials de Catalunya (IEEC), E-08034 Barcelona, Spain
    \and
    Departamento de F\'isica Matem\'atica, Instituto de F\'{\i}sica, Universidade de S\~ao Paulo, Rua do Mat\~ao 1371, CEP 05508-090, S\~ao Paulo, Brazil
    \and
    Instituto de Astrof\'isica de Andaluc\'ia (CSIC), P.O. Box 3004, 18080 Granada, Spain
    \and
    Aix Marseille Univ, CNRS, CNES, LAM, Marseille, France
    \and
    Donostia International Physics Center, Paseo Manuel de Lardizabal 4, E-20018 Donostia-San Sebastian, Spain
    \and
    Ikerbasque, Basque Foundation for Science, E-48013 Bilbao, Spain
    \and
    Department of Astronomy, University of Illinois at Urbana-Champaign, Urbana, IL 61801, USA
    \and
    Observatório Nacional, Rua General José Cristino, 77, São Cristóvão, 20921-400, Rio de Janeiro, RJ, Brazil
    \and 
    Instituto de Astrofísica de Andalucía (CSIC)
    \and
    Instituto de Física, Universidade Federal da Bahia, 40210-340, Salvador, BA, Brazil
    \and
    Centro de Estudios de Física del Cosmos de Aragón (CEFCA), Unidad Asociada al CSIC, Plaza San Juan, 1, E-44001, Teruel, Spain
    \and
    Centro de Estudios de Física del Cosmos de Aragón (CEFCA), Plaza San Juan, 1, E-44001, Teruel, Spain
    \and
    Núcleo de Astrofísica e Cosmologia \& Departamento de Física, Universidade Federal do Espírito Santo, 29075-910, Vitória, ES, Brazil
    \and 
    INAF -- Osservatorio Astronomico di Trieste, via Tiepolo 11, 34131 Trieste, Italy
    \and
    IFPU -- Institute for Fundamental Physics of the Universe, via Beirut 2, 34151, Trieste, Italy
    \and 
    Departamento de Astronomia, Instituto de Astronomia, Geofísica e Ciências Atmosféricas, Universidade de São Paulo, São Paulo, Brazil
    \and
    Instruments4, 4121 Pembury Place, La Canada Flintridge, CA 91011, U.S.A
    }
    
 
\abstract
{}
{Quasar catalogues from photometric data are used in a variety of applications including targetting for spectroscopic follow-up, measurements of supermassive black hole masses, Baryon Acoustic Oscillations or non-Gaussianities. Here, we present a list of quasar candidates including photometric redshift estimates from the miniJPAS Data Release constructed using \squeze{}. miniJPAS is a small proof-of-concept survey covering 1 $deg^2$ with the full J-PAS filter system, consisting of $54$ narrow filters and $2$ broader filters covering the entire optical wavelength range.}
{This work is based on machine-learning classification of photometric data of quasar candidates using \squeze{}. It has the advantage that its classification procedure can be explained to some extent, making it less of a `black box' when compared with other classifiers. Another key advantage is that the use of user-defined metrics means the user has more control over the classification.
While \squeze{} was designed for spectroscopic data, here we adapt it for multi-band photometric data, i.e. we treat multiple narrow-band filters as very low-resolution spectra.
We train our models using specialized mocks from \cite{Queiroz+2022}.  We estimate our redshift precision using the normalized median absolute deviation, $\sigma_{\rm NMAD}$ applied to our test sample.}
{Our test sample returns an $f_{1}$ score (effectively the purity and completeness) of 0.49 for high-z quasars (with $z\geq2.1$) down to magnitude $r=24.3$ and 0.24 for low-z quasars (with $z<2.1$), also down to magnitude $r=24.3$. For high-z quasars, this goes up to 0.9 for magnitudes $r<21.0$. We present two catalogues of quasar candidates including redshift estimates: 301 from point-like sources and 1049 when also including extended sources. We discuss the impact of including extended sources in our predictions (they are not included in the mocks), as well as the impact of changing the noise model of the mocks. We also give an explanation of \squeze{} reasoning. Our estimates for the redshift precision using the test sample indicate a $\sigma_{\rm NMAD}=0.92\%$ for the entire sample, reduced to 0.81\% for $r<22.5$ and 0.74\% for $r<21.3$. Spectroscopic follow-up of the candidates is required in order to confirm the validity of our findings.
}
{}

\date{Received XXX, 2023; accepted YYYY, 2023}
\keywords{quasars: general -- methods: data analysis -- techniques: photometric -- cosmology: observations}

\maketitle



\section{Introduction}

In recent years we have seen the appearance of several very large spectroscopic surveys, including hundreds of thousands of spectra.
Among these, dedicated cosmology programs, such as the Baryon Oscillation Spectroscopic Survey \citep[BOSS]{Dawson+2013}, part of the third extension of the Sloan Digital Sky Survey \citep[SDSS-III]{Eisenstein+2011}, where the spectra of quasars play a key role in constraining cosmological parameters. Indeed, BOSS quasars were used in the first Baryon Acoustic Oscillations (BAO) measurement using the Lyman $\alpha$ (Ly$\alpha$) forest auto-correlation \citep{Busca+2013, Slosar+2013, Kirkby+2013} and its cross-correlation with quasars \citep{Font-Ribera+2013}.

These Ly$\alpha$ BAO measurements were refined in the extended BOSS Survey \citep[eBOSS]{Dawson+2016}, part of SDSS-IV \citep{Blanton+2017}, leading to their measurements on the 16 Data Release (DR16) by \cite{duMasdesBourboux+2020}. Quasars in eBOSS were also used to perform BAO clustering analysis \citep{Hou+2021, Neveux+2020}. Both analyses were included in the final cosmological results from eBOSS \citep{Alam+2021}.

These large spectroscopic surveys use multi-object spectroscopy to achieve such a large number of objects in a reasonable amount of time. In practice, this requires the identification of quasars (or any other object of interest) in photometric surveys, so that we know where to place the optical fibres. The same is true for the next generation of surveys aiming to construct a large spectroscopic quasar sample. The next generation of surveys has already started collecting data. 

Among them, we find the Dark Energy Spectroscopic Instrument \citep[DESI][]{Desi+2016a, Desi+2016b}, which uses targeting data from the DESI Legacy Survey programs \citep{Dey+2019}, and the WEAVE-QSO Survey \citep{Pieri+2016}, part of the William Herschel Telescope Enhanced Area Velocity Explorer Collaboration \citep[WEAVE]{Dalton+2016}, which will use targeting data from the Javalambre Physics of the Accelerating Universe Astrophysical Survey \citep[J-PAS][]{Benitez+2014}. They will observe an unparalleled large sample of quasars.

While being a photometric survey, J-PAS has the particularity of using many narrow-band filters spaced every $\sim 100\angs{}$ producing pseudo-spectra (or {\it j-spectra}) of the observed objects. 
Therefore in addition to providing a target sample for WEAVE-QSO, J-PAS promises to deliver a sample of sufficient quality to enable various quasar analyses from J-PAS data alone \citep[e.g.][]{Abramo+2012}.
J-PAS is currently undergoing commissioning but they have released the data taken with their pathfinder camera, the miniJPAS data release \citep{Bonoli+2020}. 

Up until now, we have discussed the use of photometric quasar catalogues as targeting catalogues in large spectroscopic surveys as this is our main motivation to build this catalogue. However, we note that photometric quasar catalogues are also interesting in their own right. There are plenty of uses for quasars in photometric surveys, including measuring supermassive black hole masses \citep[e.g][]{Chaves-Montero+2022}, Baryon Acoustic Oscillations \citep[e.g.][]{Abramo+2012} and non-Gaussianity \citep[e.g][]{Leistedt+2014}.

Previous papers on this series \citep{Rodrigues+2023, Martinez-Solaeche+2023} introduce different flavours of machine-learning algorithms to construct a catalogue from miniJPAS data. Here, we present the classification of \squeze{} \citep{Perez-Rafols+2020b}. \squeze{} is a machine-learning code designed to identify spectra of quasars, but that can be extended to using the j-spectra from J-PAS \citep{Perez-Rafols+2020a}. The particularity of \squeze{} is that it does not only perform the classification of quasars but also provides photometric redshift estimates. 

Having redshift estimates is a key feature as they are essential if this catalogue is to be used, for example, to measure BAO \citep{Abramo+2012}. In general, previous efforts to obtain photometric quasar redshift estimates use quasar templates. Examples of this are \cite{Wolf+2003} on COMBO-17 data, \cite{Salvato+2009, Salvato+2011} on COSMOS data, \cite{Matute+2012, Chaves-Montero+2017} on ALHAMBRA data or \cite{MendesdeOliveira+2019} on S-PLUS data. Other surveys like J-PLUS \citep{Cenarro+2019, Spinoso+2020} detected the position of the Lyman $\alpha$ emission line to estimate the redshifts of Lyman $\alpha$ emitters. However, by analysing a single emission line they are susceptible to interlopers. Also, even if the detected line is indeed Lyman $\alpha$ emission, they cannot distinguish between different types of Lyman $\alpha$ emitters (i.e. star-forming galaxies and QSOs). Here, we use an alternative method that mimics the visual inspection of quasar spectra by quasar experts. Searches are performed using \squeze{} by finding multiple emission lines, their relative wavelengths and strengths. In doing so, we simultaneously provide quasar identification and redshift estimation.

We start by describing the data we use in Section~\ref{sec:data}. Then, we describe \squeze{} behaviour and its particularities when running it on J-PAS j-spectra in Section~\ref{sec:squeze} and present the results on synthetic data (mocks) in Section~\ref{sec:performance}. Then we present our catalogues of quasar candidates in Sections~\ref{sec:quasar_cat}, and discuss our findings in \ref{sec:discussion}. Finally, we summarize our conclusions in Section~\ref{sec:summary}.

\section{miniJPAS data \& mocks}\label{sec:data}
In this Section, we describe the datasets used in this work. The number of objects in each sample is summarized in Table~\ref{tab:samples}. Where known, we also give the number of quasars, galaxies and stars separately. We now describe each of these samples.

\begin{table*}
    \centering
    \caption{Summary of the samples used in this work.}
    \label{tab:samples}
    \begin{tabular}{llcccc}
        \toprule
        type & sample & objects & star & galaxy & quasar \\
        \midrule
        \multirow{4}{*}{real data} 
        & all & 40,805 & $\dots$ & $\dots$ & $\dots$ \\
        & point-like & 10,282 & $\dots$ & $\dots$ & $\dots$\\
        & SDSS cross-match & 272 & 115 & 40 & 117\\
        & SDSS cross-match \& point-like & 254 & 27 & 40 & 113\\
        \midrule
        \multirow{4}{*}{mocks} 
        & training & 297,959 & 99,931 & 99,109 & 98,919\\
        & validation & 29,783 & 9,991 & 9,901 & 9,891 \\
        & test & 29,779 & 9,995 & 9,897 & 9,887 \\
        & test 1 deg$^{2}$ & 9,036 & 2,187 & 6,347 & 502 \\

        \bottomrule
    \end{tabular}
    \\
    \vspace{0.1cm}
    The first column specifies the type of sample (mocks or real data). The second and third columns give the name of the sample and the number of objects it contains. Where available, in the following columns we also provide the number of stars, galaxies, and quasars separately.
\end{table*}

In this work, we use data from the First Data Release of J-PAS, also known as the miniJPAS survey \citep{Bonoli+2020}. The miniJPAS survey is a photometric survey using 56 filters, of which 54 are narrow band filters with FWHM $\sim 140$\AA{}  and 2 are broader filters extending to the UV and the near-infrared. These 56 filters are complemented by the $u$, $g$, $r$ and $i$ SDSS broadband filters. The survey covers $\sim1{\rm deg}^{2}$ on the AEGIS field. 

For this work, we use the sources identified using the software \texttt{SEXTRACTOR} \citep{Bertin+1996} using its dual mode. This means that the positions and sizes of the apertures used to estimate the photometry are derived from the reference filter (SDSS $r$-band). We refer the reader to \cite{Bertin+1996} and \cite{Bonoli+2020} for more detailed explanations of the software and object detection. The observations were carried out with the 2.55m T250 telescope at the Observatorio Astrofísico de Javalambre, a facility developed and operated by the Centro de Estudios de Física del Cosmos de Aragón (CEFCA), in Teruel (Spain) using the pathfinder instrument, a single CCD direct imager ($9.2k\times9.2k$, $10\mu m$ pixel) located at the centre of the T250 field of view with a pixel scale of 0.23 arcsec pix$^{-1}$, vignetted on its periphery, providing an effective FoV of 0.27 deg$^2$.

In this dual catalogue, there are a total of 64,293 identified objects\footnote{Available on \url{https://archive.cefca.es/catalogues/minijpas-pdr201912}}. A fraction of these objects are flagged as having known issues (see \citealt{Bonoli+2020} for a description of the flags). Here, we discard flagged objects to construct a clean sample of 46,440 objects. However, since high-redshift quasars are typically point-like sources, our main sample is limited to point-like sources by using the stellarity index constructed from image morphology using Extremely Randomized Trees \citep[ERT][]{Baqui+2021}. Following \cite{Queiroz+2022, Rodrigues+2023, Martinez-Solaeche+2023}, we require objects to be classified as stars (point-like sources) with a probability of at least 0.1, defined in their catalogue as $ERT\geq0.1$. In some cases, the ERT classification failed (identified as ERT=-99.0). In these cases, we used the alternative classification using the stellar-galaxy locus classifier from \cite{Lopez-Sanjuan+2019}, requiring a minimum probability of $SGLC\geq0.1$. 11,419 objects meet this point-like source criterion and constitute our point-like sample.

A small number of the objects observed in miniJPAS have spectroscopic observations from other surveys. This allows us to have a spectroscopically confirmed classification of these objects. In particular, 272 objects were observed also by SDSS, of which 117 are quasars, 40 galaxies, and 115 stars. Of these, 18 were not classified as point-like sources following the aforementioned criteria 4 are quasars, 1 is a star and 13 are galaxies.
 
In this work, apart from using miniJPAS data we also use synthetic data (mocks). This is necessary because larger data volumes with associated truth tables are needed than are currently available. The mocks we use here are based on SDSS spectra convolved with the J-PAS filters and with added noise to match miniJPAS expected signal-to-noise. More details on the mocks can be found in \cite{Queiroz+2022}. There are a total of 360,000 objects distributed between the training (300,000), validation (30,000), and test (30,000) sets. They are evenly split between quasars, galaxies, and stars. Additionally, we have a special 1 deg$^2$ test set which has the expected relative fraction for each type of object. In general, we use the mocks generated using noise model 11, since that is the noise model believed to be closest to the actual noise distribution from miniJPAS data, but we check the impact of choosing a different noise model in Appendix~\ref{sec:noise}.

Both in data and in mocks, we restrict ourselves to r-band magnitudes $17.0 < r \leq 24.3$. In the process of creating the mocks, the original spectra are rescaled to match the expected magnitude distribution. Noise is added after this rescale is done, modifying the reported values of the flux (and thus the magnitudes). This means that a few of the mocks end up with a magnitude that is fainter than 24.3. Here we discard these spectra. Overall we analyze 40,805 miniJPAS sources, of which 10,282 meet the criteria of being point-like sources. 272 sources have spectroscopic observations from SDSS, of which 254 meet the criteria of being point-like sources. \squeze{} is trained using 99,931 stars, 99,109 galaxies, and 98,919 quasars and validated using 9,991 stars, 9,901 galaxies and 9,891 quasars. The test sample contains 9,995 stars, 9,897 galaxies, and 9,887 quasars. The special 1 deg$^2$ test sample contains 2,187 stars, 6,347 galaxies, and 502 quasars.

\section{\squeze{} description and setup}\label{sec:squeze}
In this Section, we provide a brief description of \squeze{} and explain the particularities of applying it to photometric data from J-PAS. A full, detailed description of \squeze{} is given in \cite{Perez-Rafols+2020b}. \squeze{} is a quasar classifier that works in three steps. In the first step, we identify peaks in the spectra. We then assign trial redshifts to these peaks in the second step and we end by classifying these trial redshifts to discriminate between the correct and incorrect identifications. We now detail each of the steps.

\subsection{Peak identification}\label{sec:peak_finding}
The first step is peak identification. In this step, emission lines are identified in the spectra. In \squeze{}, this step is performed by a very simple peak finder: each spectrum is first smoothed and then peaks are located by finding those pixels with higher flux than the two nearby pixels. 

miniJPAS data contains j-spectra with data in 56 filters and the broad emission lines are expected to typically cover 3 filters (though some very broad quasar emission lines can span more than 3 J-PAS filters, see e.g. Figure 2 of \citealt{Chaves-Montero+2022}). 
This means that any smoothing we apply will typically decrease the signal-to-noise of the peak detection. However, using this simple peak finder we have many peaks that are purely arising from noise: cases where a filter has more flux than their neighbours. 

To solve this, we developed a new, more refined peak finder\footnote{The new peak finder is now included in the \squeze{} package.} (see Appendix~\ref{sec:peak_finder_comparison} for details of its performance compared to the original peak finder). The new peak finder works as follows. First, a power-law fit is applied to reproduce the continuum emission. Outliers to this fit, defined as the data points which are off the fit by more than $N$ sigmas, are discarded, and the process is repeated until convergence is reached i.e. when no data points are discarded in an iteration. Here, $N$ is the minimum significance to detect outliers, and we choose $N=2$ as our fiducial choice (see Section~\ref{sec:significance}). Upon fit convergence, the outliers below the model are discarded and the outliers above the model are kept as emission peaks. Contiguous peaks (i.e. with pixel number $i$, $i+1$, $i+2$, ...) are compressed into a single peak by performing a weighted average of their wavelengths. The weights are defined by the significance of the outliers. Sometimes, too many pixels are discarded as outliers and the power-law fit fails. In those cases, no emission peaks are reported and the spectra are discarded. The overall performance of \squeze{} is improved when the new peak finder is used (see Appendix~\ref{sec:peak_finder_comparison}).

\subsection{Trial redshifts}\label{sec:trial_z}
\begin{figure}
    \centering
    \includegraphics[width=0.35\textwidth]{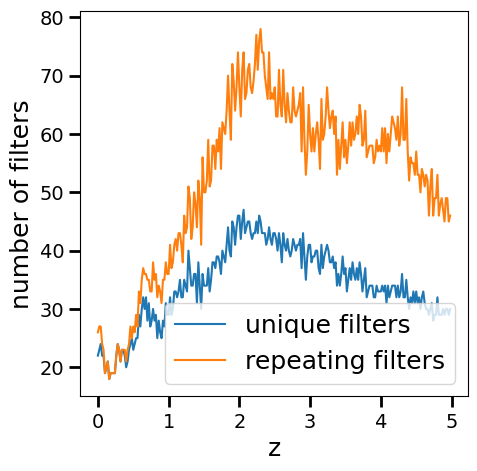}
    \caption{Number of J-PAS filters used to compute the metrics as a function of redshift. The blue line shows the number of unique filters used in the computation of the metrics. Given that some bands are overlapping, the total number of filters used is generally larger as shown by the orange line.}
    \label{fig:num_filters}
\end{figure}

Once the peaks have been identified, a list of trial redshifts is generated. For each peak of each spectrum, a trial redshift, $z_{\rm try}$, is computed assuming that the peak corresponds to the \lya{}, \civ{}, \ciii{}, \mgii{}, \ha{}, and \hb{} emission lines. Negative trial redshifts are immediately discarded. Line metrics are computed for each of the remaining trial redshifts as described in Equations 1 to 3 of \cite{Perez-Rafols+2020b}. These metrics describe the amplitude of the line, its significance, and the slope at the base of the line. For each trial redshift, we compute the metrics for 17 bands (see Table~\ref{tab:line_confusion_new_lines}) corresponding to the predicted position of quasar emission lines and other relevant features (see Appendix~\ref{sec:new_lines} for details). These bands are defined at the potential quasar rest frame, and therefore the spectral coverage of these bands will change as a function of redshift. Figure~\ref{fig:num_filters} shows this evolution as the number of filters used in line metrics as a function of redshift. 

The wavelength bands used to compute line metrics were designed and optimized for spectra from BOSS, with a resolution of $\sim1\angs$, while here the resolution is $140\angs$ and the separation between filter centres is $100\angs$. We have, therefore, explored whether the size of the bands impacts the classification performance. The boundaries of these bands have been tuned to more reliably separate peak from side-band in data of this resolution. Furthermore, given our limited use of the filters available we have added bands to allow the machine learning algorithm (see Section~\ref{sec:classification}) access to information on the absence of emission lines as well as the presence of them. These "flat emission line" metrics are placed at wavelengths where redshift confusion leads to an emission line arising where one should not occur. The desired lack of emission line signal can hence be part of the random forest classification. See Section~\ref{sec:new_lines} for more details on the tests that led to our selection of this `wide+extra' set of wavelength bands. We note that even with our more inclusive approach, only a relatively small number of the filters (out of a total of 60 available) are used to compute the metrics and, thus, identify the quasars (see the left panel in Figure~\ref{fig:new_lines_filters_used}).

\subsection{Classification}\label{sec:classification}
Once we have a list of trial redshifts and associated metrics, they are fed to the random forest classifiers. In training mode, trial redshifts are flagged as correct peaks if the spectra are that of a quasar with trial redshift at most $\Delta z=0.10$ away from the true redshift. In \cite{Perez-Rafols+2020b}, they use a larger value of 0.15 for this criterion, but we get better redshift errors (without a decrease in performance) by using a tighter constraint (see Section~\ref{sec:z_tolerance}). 

In \cite{Perez-Rafols+2020b}, two different classifiers were used, one for high redshift quasars and another one for low redshift quasars. The split in redshift was performed at $z=2.1$ since this is where the \lya{} emission line enters the spectra. They argue that a single classifier could be used but that they observe better performance when splitting by redshift. The reason for this is that high redshift quasars have more emission lines compared to low redshift quasars. We checked that this statement is also valid for our dataset (see Section~\ref{sec:new_rf}) and decided to also use two random forests. In \squeze{} default choices, only the metrics are passed to the random forest classifiers. However, we note that by also passing the trial redshift and the r-band magnitude we obtain slightly better results (see Section~\ref{sec:add_cols}). Thus, we adopt this change of the default settings.

\begin{figure}
    \centering
    \includegraphics[width=0.8\columnwidth]{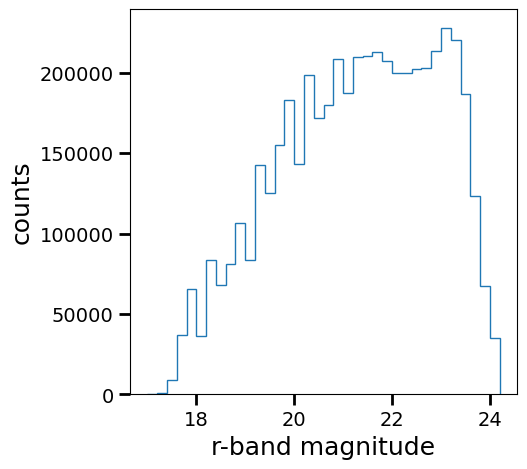}
    \caption{Magnitude distribution of the trial redshifts in the training sample. This distribution does not match the distribution of objects as there are typically a few trial redshifts per object.}
    \label{fig:train_mag_dist}
\end{figure}

The final stage in the classification is to select, for each spectrum, the trial redshift with the highest probability. At this point, it is worth noting that it is more convenient to separate quasars by their observed r-band magnitude as the faint quasars dominate the training set. This can be seen in Figure~\ref{fig:train_mag_dist}, where we show the magnitude distribution of the list of trial redshifts. Here, we note that this distribution is different from the distribution of objects, as each object will typically have a few trial redshifts. As explained in detail in Section~\ref{sec:mag_bins}, we run \squeze{} in four magnitude bins: $r\in\left(17.0, 20.0\right]$, $r\in\left(20.0, 22.5\right]$, $r\in\left(22.5, 23.6\right]$ and $r\in\left(23.6, 24.3\right]$.

\section{Performance}\label{sec:performance}
Here we assess the performance of \squeze{} based on the test sample results. In any sample (either from mocks or from real data) there are quasars, galaxies, and stars alike. However, we need to keep in mind that here we are interested in creating a quasar catalogue. Thus, in terms of performance, we do not need to penalize the cases where stars are classified as galaxies and vice-versa. We estimate our performance based on the correctly classified quasars. However, for a correct classification, we require not just that the object is indeed a quasar, but also that its redshift is correct. Formally, we require $\Delta z = \left|z_{\rm true} - z_{\rm try}\right| < 0.10$ (see section~\ref{sec:z_tolerance}) as this is enough to ensure that we are not suffering line confusion (i.e. finding a true emission line but failing to label it correctly). We note, however, that the actual redshift precision is typically better (see below). 

We define purity $p$ as the number of true quasars (at the correct redshift) in the catalogue over the total number of sources classified as quasars, and completeness $c$ as the number of true quasars in the catalogue (again, at the correct redshift) over the total number of true quasars in the sample analysed. For each of the classifications, we also have the confidence of the classification, given by the fraction of decision trees that agree with that classification. To some extent, we can tune the purity and completeness of the sample by applying some cuts on this confidence of classification.

A higher confidence requirement will result in a purer but less complete sample. Similarly, a lower confidence requirement will result in a more complete, but less pure, sample. Even though the choice of a confidence threshold can be tuned for specific analysis, a common general-purpose choice is to balance purity and completeness. An optimised balance can be found by maximizing the $f_{1}$ score, defined as:
\begin{equation}
    \label{eq:f1}
    f_{1} = \frac{2pc}{p+c} ~.
\end{equation}

We note that with this definition, performance is expected to be worse than the other classifiers presented in the companion papers \citep{Rodrigues+2023, Martinez-Solaeche+2023}. Part of the reason
for this is that they have a more relaxed criterion to determine good classifications. Since they are not measuring redshifts, they require the quasars to be correctly classified as high redshift quasars ($z\geq 2.1$) or low redshift quasars ($z< 2.1$). We also adopt this criterion to make a more direct comparison. We denote this criterion as $f_{1}^{*}$.

\subsection{Test sample}\label{sec:performance_test}
\begin{figure*}
    \centering
    \includegraphics[width=0.7\textwidth]{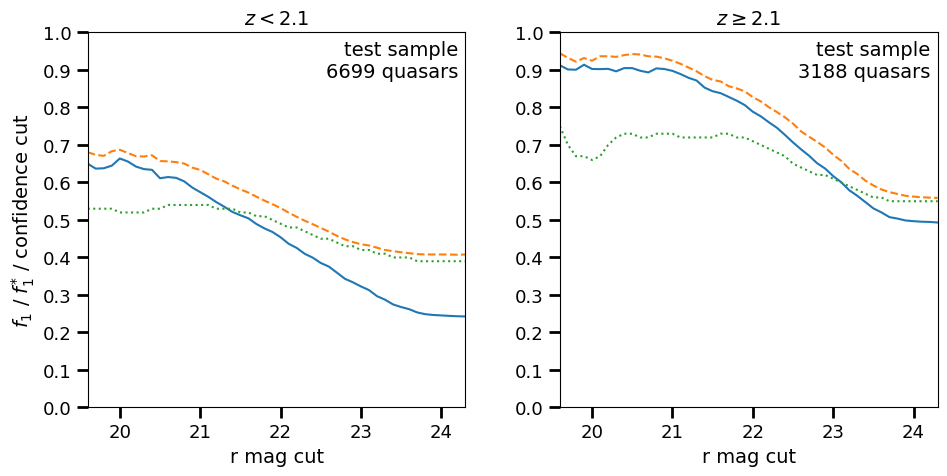}
    \includegraphics[width=0.7\textwidth]{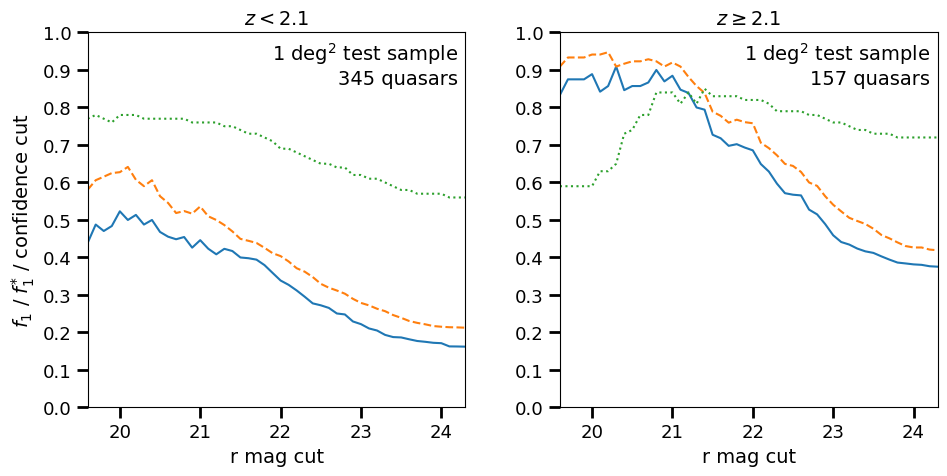}
    \includegraphics[width=0.7\textwidth]{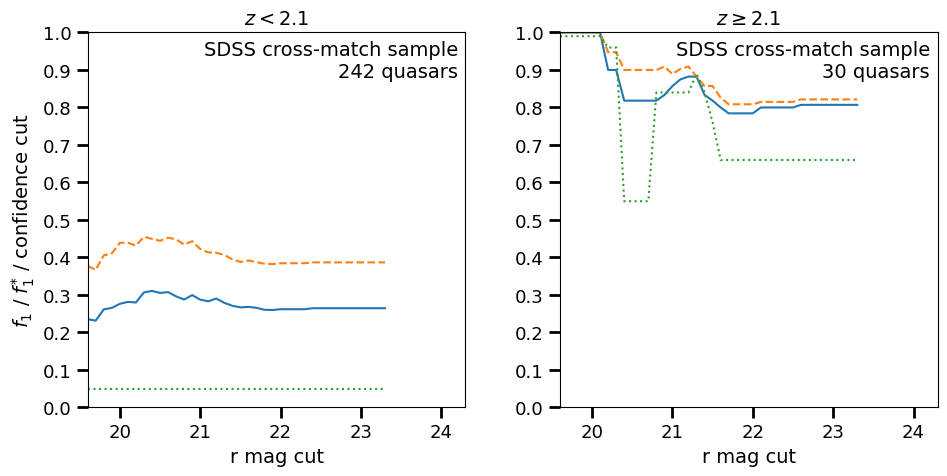}
    \caption{Performance as a function of limiting magnitude. All objects brighter than the magnitude cut in the $r$ band are considered to compute the $f_1$ score. Blue solid lines show the $f_1$ score as defined in Equation~\ref{eq:f1} and the orange dashed lines show the more relaxed statistic $f_{1}^{*}$ (see text for details). Green dotted lines show the confidence threshold used to compute the $f_{1}$ score. From top to bottom results for the test sample (Section~\ref{sec:performance_test}), the 1 deg$^{2}$ test sample (Section~\ref{sec:performance_1deg2_test}), and the SDSS cross-match sample (Section~\ref{sec:performance_sdss}). We note that in the bottom panels, the lines stop at magnitude 23.4 the sample does not have fainter objects.}
    \label{fig:f1_vs_mag}
\end{figure*}

The top panels in Figure~\ref{fig:f1_vs_mag} show the performance as a function of limiting magnitude. Blue solid lines correspond to the $f_{1}$ score, whereas the orange dashed lines show the more relaxed criterion $f_{1}^{*}$. For each limiting magnitude, we perform a cut in the confidence threshold of the classification such that the $f_1$ score is maximized (green dotted lines). As expected, the performance drops as fainter objects are added to the sample. This is because fainter objects are more difficult to classify as they are noisier and have a larger number of filters with non-detections. The $f_1$ score including all objects down to $r=24.3$ is 0.49 (with a confidence threshold of 0.55) for high-z quasars and 0.24 for low-z quasars (with a confidence threshold of 0.39). The values of $f_{1}^{*}$ are higher than those of $f_{1}$, as expected. Including all objects down to $r=24.3$ its values are 0.56 for high-z quasars (with a confidence threshold of 0.58) and 0.41 for low-z quasars (with a confidence threshold of 0.32). 

\begin{figure}
    \centering
    \includegraphics[width=\columnwidth]{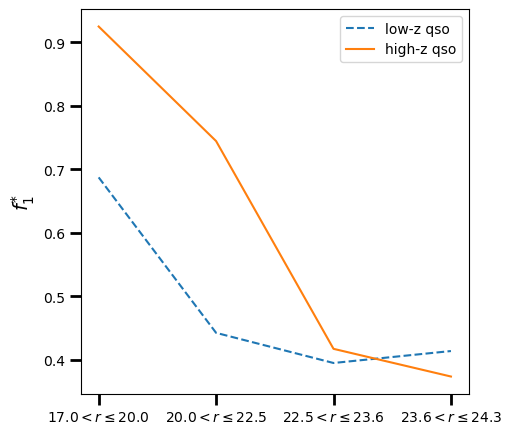}
    \caption{$f_{1}^{*}$ measured using the same bins as \protect\cite{Rodrigues+2023, Martinez-Solaeche+2023}. The solid orange line shows the $f_1^{*}$ score for high-z quasars and the dashed blue line for low-z quasars.}
    \label{fig:f1_mag_bins}
\end{figure}

The comparison of these results with those obtained by the algorithms from \cite{Rodrigues+2023, Martinez-Solaeche+2023} is not straightforward. They report the averaged $f_{1}$ score including the $f_{1}$ for high-z quasars, low-z quasars, galaxies and stars. Here, by using $f_{1}^{*}$, we can only compute equivalent quantities for high-z quasars and low-z quasars. As such, only a qualitative comparison is possible. Even so, we provide our measurement of $f_{1}^{*}$ using the same magnitude bins in Figure~\ref{fig:f1_mag_bins}. This Figure should be compared with the top panel of Figure 4 of \cite{Rodrigues+2023} and with Figure 1 of \cite{Martinez-Solaeche+2023}. 

We can see that \squeze{} performance is qualitatively higher than the RF from \cite{Rodrigues+2023}. It has a qualitatively similar performance compared to LGMB and CNN1 (without errors), also from \cite{Rodrigues+2023} and is qualitatively lower than CNN1, CNN2 from \cite{Rodrigues+2023} and the classifiers from \cite{Martinez-Solaeche+2023}. However, this is expected as our method tackles the harder problem of solving both the redshift estimation and the classification problems.

\begin{figure*}
    \centering
    \includegraphics[width=\textwidth]{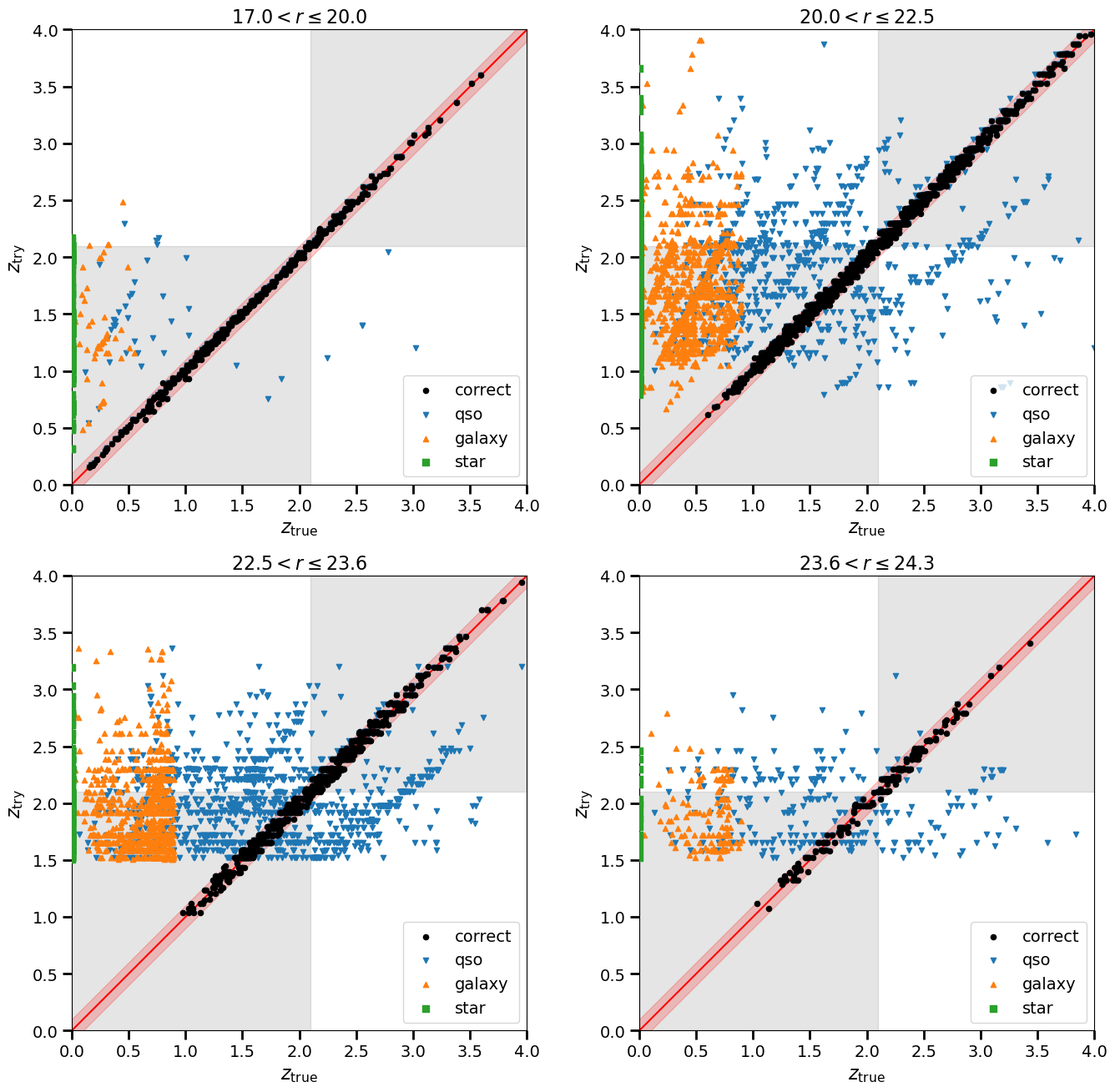}
    \caption{\squeze{} trial redshift, $z_{\rm try}$, vs true redshift, $z_{\rm true}$, for the test sample. Black dots indicate correct classifications. Blue down-pointing triangles, orange up-pointing triangles, and green squares indicate the quasar, galactic, and stellar contaminants respectively. The red solid line shows the perfect classification line and the red stripe show the redshift offset tolerance ($\Delta z = 0.10$). Grey squares indicate the area where quasar contaminants are deemed as correct in the relaxed classification scheme (see text for details). From top to bottom and left to right, panels show the data split in four different magnitude bins: $17 < r \leq 20$, $20 < r \leq 22.5$, $22.5 < r \leq 23.6$, and $23.6 < r \leq 24.3$. Note that stellar contaminants (green dots) are always found at $z_{\rm true}=0$. }
    \label{fig:line_confusion}

\end{figure*}

We now move to analysing the contaminants of our sample. To do so, we split it into four different magnitude bins: $17 < r \leq 20$ (bin 1), $20 < r \leq 22.5$ (bin 2), $22.5 < r \leq 23.6$ (bin 3), and $23.6 < r \leq 24.3$ (bin 4). For each of the bins, we plot the predicted redshift, $z_{\rm try}$, against the true redshift, $z_{\rm true}$, to study the contaminants. We include only quasars with confidences greater than 0.39 when $z_{\rm try}<2.1$, and 0.55 when $z_{\rm try}\geq2.1$, corresponding to the confidence thresholds mentioned above. Figure~\ref{fig:line_confusion} shows galaxies, stars, and quasar contaminants as orange up-pointing triangles, green squares, and blue down-pointing triangles, respectively. Black dots show correct classifications, i.e., those that fulfil the criteria of $\Delta z = \left|z_{\rm true} - z_{\rm try}\right| < 0.10$ (red band). Grey bands also signal the areas where quasar contaminants (blue down-pointing triangles) are correctly classified in the relaxed classification criterion, i.e., without a redshift requirement.

Bin 1 behaves as expected. Correct classifications are found very close to the red line, showing that the redshift precision is significantly better than the required value of 0.10 (indicated in the plot by the red stripe). Quasar contaminants are rare and follow straight lines showing that there is some degree of line confusion, i.e. we correctly find a quasar emission line but we fail to identify the line responsible for the emission (thus the redshift error). Galactic contaminants also follow the same straight lines as quasars suggesting that these galaxies might contain Active Galactic Nuclei (AGNs) with broad emissions lines or --more likely-- star-formation emission lines (\ha{}, \hb{}, \oiii{}, \oii{}) that are misidentified as QSO emission lines \citep{Chaves-Montero+2017}. Stellar contaminants are distributed at trial redshifts lower than 2.1. This indicates that only the low-z classifier is adding stellar contaminants.

This simple picture starts to break as we go to bin 2.  We see two effects. First, while we still see clear line confusion, we also see some quasar and galactic contaminants that are no longer distributed along these lines, indicating that we are no longer able to always distinguish real emission line peaks from noise peaks. Apart from this, we see by eye that the redshift precision of the correct classifications is significantly worse (see below for a more quantitative statement). This suggests that the chosen redshift tolerance was too large. We discuss this further in Section~\ref{sec:z_tolerance}, where we conclude that this is not the case. 

This issue is aggravated as we go to bin 3. Now we are not able to see the confusion lines as clearly as before (though some can still be seen). This suggests that our ability to distinguish real peaks from noise peaks starts to break somewhere around magnitude $r\sim22.5$ (see also Section~\ref{sec:peak_finder_comparison}). We note that in this bin we see an apparent cut of the contaminants in the redshift at $z=1.5$. Below this redshift, the \ion{C}{IV} line is not observable in our spectral coverage. Together with the \lya{} line, they are the stronger lines. Thus, the confidence in classification is generally lower whenever they are not present. In practice, this means that many of the trial redshifts below this redshift either do not meet the minimum required classification confidence, or else other trial redshifts for the same quasar are preferred. 

Finally, for bin 4, there is a strong decrease in the number of contaminants. There are two reasons behind this. First, we are at the faint end of our sample and therefore the number of objects decreases compared to bin 3. Second, the spectra are so noisy that 
we obtain very few confident classifications.

Overall, a significant fraction of the redshift confusion is causing high-z quasars (with $z\geq2.1$) to be classified as low-z quasars. This can be seen in the upper left quadrants in Figure~\ref{fig:line_confusion}. In lower numbers, the same occurs in the opposite direction (lower right quadrants ). This can explain the drop in performance compared to the results from \cite{Rodrigues+2023, Martinez-Solaeche+2023} even when we consider the same relaxed criteria $f_{1}^{*}$. However, the fact that there is more than one trial redshift per quasar indicates that if the high/low redshift classification could be fixed by these other algorithms, \squeze{} can still be used to provide a redshift estimate (see Pérez-Ràfols et al. In prep.).

\begin{figure}
    \centering
    \includegraphics[width=\columnwidth]{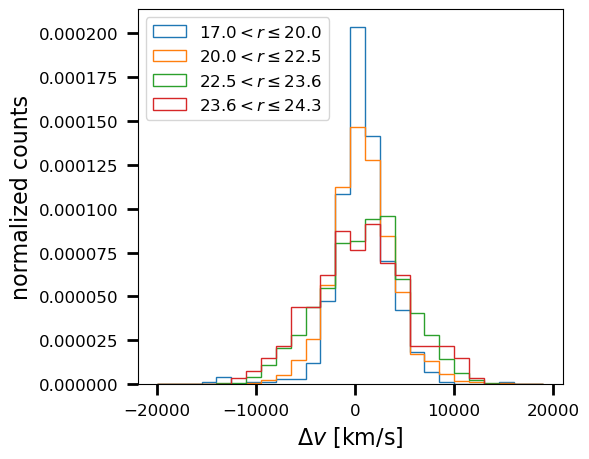}
    \caption{Distribution of $\Delta z = \left|z_{\rm true} - z_{\rm try}\right| < 0.15$ for the test sample for four magnitude bins: $17 < r \leq 20$, $20 < r \leq 22.5$, $22.5 < r \leq 23.6$, and $23.6 < r \leq 24.3$. }
    \label{fig:z_dist}
\end{figure}

We now turn our attention to the redshift precision. As explained in Section~\ref{sec:squeze}, we formally require a precision of 0.10 but the performance is expected to be much better. Figure~\ref{fig:line_confusion} shows that this is not the case for the fainter bins. We now quantify this statement. We take the correct classifications and measure the distribution of $\Delta z$ for the different magnitude bins. Results of this exercise are shown in Figure~\ref{fig:z_dist} and in the first block of Table~\ref{tab:z_dist_test}. Indeed, the redshift error increases as we go to fainter magnitudes. In fact, our bright bin (bin 1, with $17.0 < r \leq 20.0$) has a typical redshift error of $\sim 2800{\rm km/s}$, which is less than two-thirds of the typical error in our faint bin (bin 4, $\sim 4700{\rm km/s}$). We also see that there is no significant bias in our measurement of the redshift (the mean offset is an order of magnitude smaller than the typical error). 

\begin{table*}
    \centering
    \caption{Statistics of the redshift precision.}
    \label{tab:z_dist_test}
\begin{tabular}{cccccc}
\toprule
             sample & mag bin &  $\overline{\Delta v}$ (km/s) &  $\sigma_{\Delta v}$ (km/s) &  $\sigma_{\rm NMAD}$ (per cent) &  $N$ \\
\midrule
\multirow{4}{*}{test}
& $17.0 < r \leq 20.0$ &                        704.69 &                    2,833.20 &                    0.71 &  504 \\
& $20.0 < r \leq 22.5$ &                        809.09 &                    3,083.64 &                    0.94 & 2265 \\
& $22.5 < r \leq 23.6$ &                        859.76 &                    4,333.32 &                    1.51 &  983 \\
& $23.6 < r \leq 24.3$ &                        372.16 &                    4,702.21 &                    1.56 &  183 \\
\midrule
\multirow{4}{*}{1 deg$^2$ test}
& $17.0 < r \leq 20.0$ &                      1,558.23 &                    2,753.08 &                    0.75 &   35 \\
& $20.0 < r \leq 22.5$ &                        787.62 &                    3,212.02 &                    0.86 &  104 \\
& $22.5 < r \leq 23.6$ &                      2,121.36 &                    4,577.93 &                    1.80 &   58 \\
& $23.6 < r \leq 24.3$ &                      2,954.34 &                    5,542.67 &                    1.41 &    9 \\
\midrule
\multirow{3}{*}{SDSS cross-match sample}
& $17.0 < r \leq 20.0$ &                      1,667.23 &                    2,361.88 &                    0.73 &   22 \\
& $20.0 < r \leq 22.5$ &                      1,263.28 &                    2,625.00 &                    0.88 &   61 \\
& $22.5 < r \leq 23.6$ &                      1,084.38 &                         NaN &                    0.53 &    1 \\
\bottomrule
\end{tabular}
    \\
    \vspace{0.1cm}
    Mean redshift offset ($\overline{\Delta v}$), dispersion measured ($\sigma_{\Delta v}$), normalised median absolute deviation ($\sigma_{\rm NMAD}$) and number of correctly classified quasars ($N$) in the samples test, 1 deg$^2$ test, and SDSS cross-match. The mean offset indicates potential biases of our redshift estimate and the dispersion indicates our typical redshift error. The number of objects indicates how reliable are the measured statistics.
\end{table*}

We also provide in Table~\ref{tab:z_dist_test} the normalised median absolute deviation, $\sigma_{\rm NMAD}$, defined by \cite{Hoaglin+1983} as
\begin{equation}
    \label{eq:NMAD}
    \sigma_{\rm NMAD} = 1.48 \times {\rm median} \left(\frac{\left|z_{\rm try} - z_{\rm true}\right|}{1+z_{\rm true}}\right) ~.
\end{equation}
This quantity is less sensitive to redshift outliers than the standard deviation. Nevertheless, we observe the same trend here as we do for the standard deviation.

\subsection{1 deg\texorpdfstring{$^{2}$}{2} test sample}\label{sec:performance_1deg2_test}
Now we focus on the special 1 deg$^{2}$ test sample, which differs from the normal test sample in the relative number of objects. This sample has the number of quasars, stars, and galaxies expected in a square degree on the sky. Thus, in proportion, the number of quasars is significantly smaller. The middle panel of Figure~\ref{fig:f1_vs_mag} shows the performance as a function of limiting magnitude. We see a similar trend as for the test sample. The $f_{1}$ score including all objects down to $r=24.3$ is 0.38 (with a confidence threshold of 0.72) for high-z quasars and 0.16 for low-z quasars (with a confidence threshold of 0.56). The values of $f_{1}^{*}$ including all objects down to $r=24.3$ are 0.42 (with a confidence threshold of 0.83) for high-z quasars and 0.21 for low-z quasars (with a confidence threshold of 0.68). This decrease compared to the test sample is expected as we now have a larger fraction of contaminants. 

\begin{figure}
    \centering
    \includegraphics[width=\columnwidth]{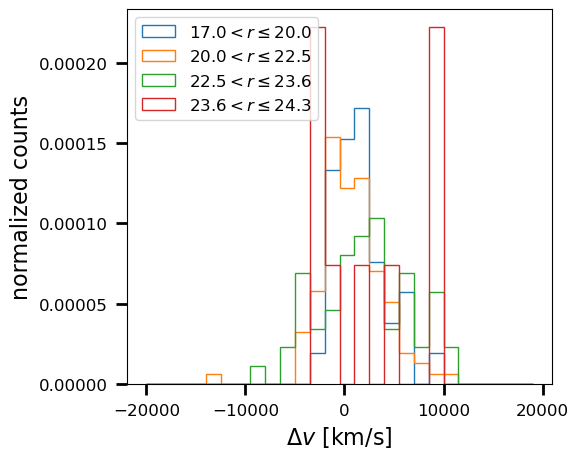}
    \caption{Same as Figure~\ref{fig:z_dist} but for the 1 deg$^{2}$ test sample.}
    \label{fig:z_dist_1deg2}
\end{figure}

We now analyse the redshift precision in this sample. As for the test sample, we compute the distribution of $\Delta z$ for different magnitude bins. The results are shown in Figure~\ref{fig:z_dist_1deg2} and tabulated in the second block of  Table~\ref{tab:z_dist_test}. The distribution of redshift errors in bins 1, 2 and 3 are similar to those of the test sample. For the faint bin, we clearly do not have enough statistics to say anything meaningful. 

\subsection{SDSS cross-match sample}\label{sec:performance_sdss}
More interesting than the performance on mock data is the performance on real data. However, we are limited in this assessment by the lack of a large sample with an available truth table. The only samples with reliable spectroscopic confirmation of the object classes are the SDSS cross-match samples (including and excluding extended objects). We note that both samples are very small (see Table~\ref{tab:samples}) and that they are biased as they only contain the brightest objects. 
Even though they are not included in the mocks, we include the 18 objects not classified as point-like sources in the performance assessment to have a sample as large as possible. We note that the results stay the same including or not including these 18 objects. 

\begin{figure}
    \centering
    \includegraphics[width=\columnwidth]{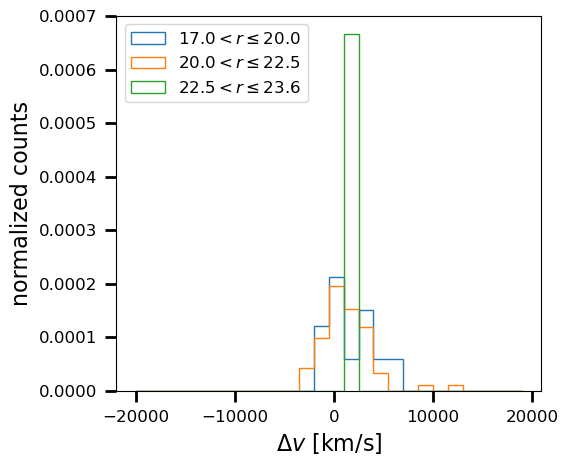}
    \caption{Same as Figure~\ref{fig:z_dist} but for the SDSS cross-match sample.}
    \label{fig:z_dist_sdss}
\end{figure}

The performance as a function of magnitude is given in the bottom panel of Figure~\ref{fig:f1_vs_mag}. The distribution of $\Delta z$ for the classifications in Figure~\ref{fig:z_dist_sdss} and summarized in the third block of Table~\ref{tab:z_dist_test}. Due to the small size of the sample, the measured $f_1$ score distribution is much noisier than in the mock test sample. However, the results suggest a similar performance as for the 1 deg$^{2}$ test sample. The redshift distribution is also noisier.

This sample does not include any object for the faintest bin (bin 4) and only one object for bin 3. This is important as the algorithm is having more difficulties when classifying objects in these fainter bins. In order to properly assess the performance on data, we will need a larger sample of spectroscopic observations would be needed, particularly including the objects at the faint end.

\section{miniJPAS Quasar catalogue}\label{sec:quasar_cat}
Now that we assessed the performance of \squeze{}, we shift our attention to the actual catalogue. We create two different catalogues: one including only point-like sources and one including also extended sources (sample \textit{all}). These samples are described in Section~\ref{sec:data} and in Table~\ref{tab:samples}.

We run \squeze{} on these two samples and add to the catalogue those objects having classification confidence higher than the cut (green dotted lines in the top panel of Figure~\ref{fig:f1_vs_mag}). Here we use the thresholds from the test sample. Even though the number of contaminants in the data should be closer to the 1 deg$^{2}$ test sample, this sample is too small for its cuts in classification confidence to be statistically robust. Thus, we choose those of the test sample. For the final catalogue, we also drop those entries flagged as duplicated to keep only one entry per object. For some objects, no peaks were found and thus they did not enter the random forest classifiers. This occurred for 906 objects in the point-like sample and 3665 objects in the entire sample. These are dropped from our final catalogue. 

The final catalogue contains 301 quasar candidates for the point-like sample and 1049 when also including extended sources. Applying the same criteria for the 1 deg$^{2}$ test sample, we obtain a catalogue of 412 quasar candidates. These numbers should be compared to those of the point-like sample, as the mocks were built to match that sample. The similarity between the number of candidates could be suggestive of similar behaviour of \squeze{} on data and mocks. 

\begin{figure*}
    \centering
    \includegraphics[width=0.8\textwidth]{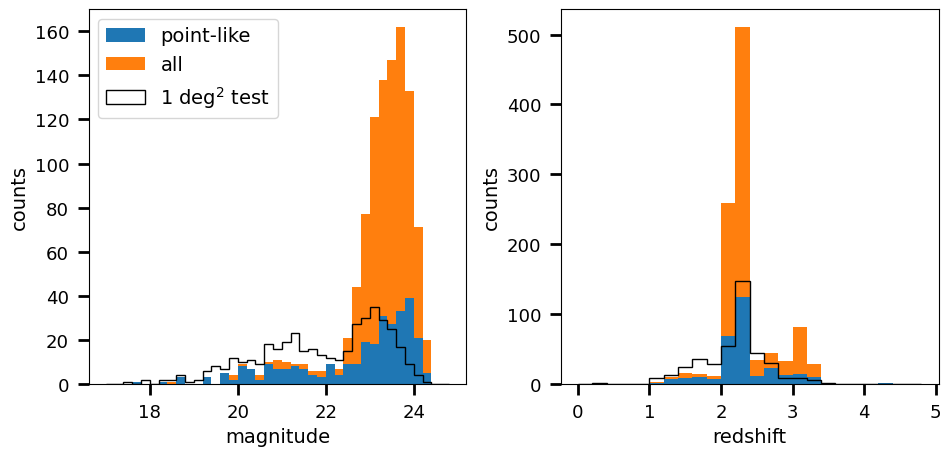}
    \caption{Solid histograms show the distribution of magnitude (left) and redshift (right) for the quasar candidates for the samples point-like and all (see Table~\ref{tab:samples}). Both distributions look similar, except for a peak at $z \sim 3$, present only when extended sources are included, suggesting the code might also work for extended sources. For comparison, empty histograms show the same distributions for the 1 deg$^{2}$ test sample. Note that the redshift used here is the best trial redshift for each candidate.}
    \label{fig:data_dist}
\end{figure*}

To dig further into this, we compare the magnitude and redshift distributions of the candidates in the point-like sample to the distributions of the candidates in the 1 deg$^{2}$ test sample (Figure~\ref{fig:data_dist}). We start with the magnitude distribution (left panel of Figure~\ref{fig:data_dist}). There is a deviation of the point-like sample towards fainter magnitudes. While small deviations are expected given the relatively small sample sizes, this could also indicate \squeze{} is performing differently in data and mocks at the faint end. Mocks were created from brighter SDSS data so it would not be surprising if a different type and/or distribution of contaminants appears at the faint end. A different population of contaminants could easily induce a different behaviour in the classifier. A spectroscopic follow-up of these sources is needed to confirm or deny this apparent discrepancy. 

The redshift distribution of the samples \textit{point-like} and \textit{all} (right panel of Figure~\ref{fig:data_dist}) are similar, except for a peak at $z\sim3$, present only when extended sources are included. This could hint that \squeze{} performance on extended sources is similar to that of point-like sources, even if it was only trained on point-like sources. There are two possible explanations for this behaviour. First, the algorithm to separate point-like sources from extended sources has a certain degree of confusion leading to some extended sources entering the point-like sample. Indeed, this seems to be happening to some degree. For instance, in the SDSS cross-matched sample we have 115 stars, of which only 27 are classified as point-like sources. This would mean that their properties are indeed included in the training sample, thus explaining the similarity between the two distributions. This is expected to happen to some degree, particularly at the faint end, where it is not always trivial to separate the actual source from the sky contribution.

Another possible explanation is that the properties of extended and point-like quasars, as seen by \squeze{}, are similar. This also makes sense as \squeze{} is focusing on the emission lines at specific spectral regions. Observing the galactic emission (and making them extended objects) would not change how \squeze{} sees the quasars. However, we note that this could change the way contaminants are seen.  Most likely, the truth lies somewhere between the two explanations, but we require a larger sample, with spectroscopic confirmation of the classifications, to ascertain this.

\section{Discussion}\label{sec:discussion}

\subsection{Comparison with previous performance estimates}\label{sec:old}
\cite{Perez-Rafols+2020a} studied the potential performance of \squeze{} in different surveys including tests for a generic narrow-band survey. In particular, they mentioned that realistic JPAS mocks, such as those we use here, would be required asses the performance of \squeze{} on JPAS data. Nevertheless, they suggested that their {\it rebin100+noise4} could be used as an initial test of this performance. Based on this they predicted the purity and completeness to be greater than 0.9. This is clearly in conflict with the results obtained here, where this statement only holds to magnitude $r < 21.1$. The simplest explanation for this discrepancy lies in the spectra used to asses this performance. To construct their {\it rebin100+noise4}, \cite{Perez-Rafols+2020a} rebinned SDSS spectra and added noise in a crude simulation of miniJPAS-like data. In this work, we assess the performance using a set of refined mocks from \cite{Queiroz+2022} that are tailored to match the observations. This discrepancy in the performance would be expected if the initial estimates from \cite{Perez-Rafols+2020a} were optimistic in the expected signal-to-noise. 

\begin{figure}
    \centering
    \includegraphics[width=\columnwidth]{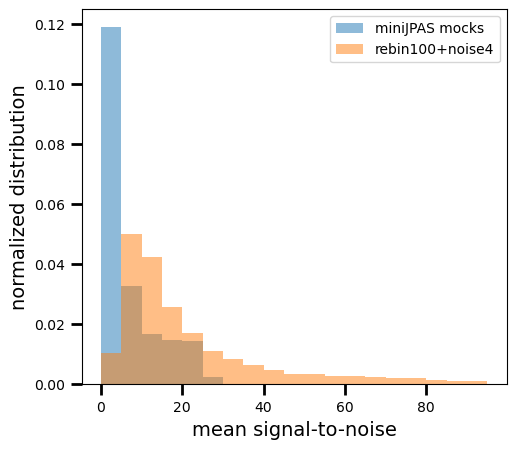}
    \caption{Normalized distributions of the mean signal-to-noise ratio for the spectra of the test sample (blue) and the rebuilt test sample (orange). We rebuild the test sample following the \cite{Perez-Rafols+2020b} prescription for building the rebin100+noise4 mocks.}
    \label{fig:snr}
\end{figure}

In order to test this, we rebuild the train and test samples but instead of using the miniJPAS mocks outlined here, we follow the \cite{Perez-Rafols+2020b} prescription for building the rebin100+noise4 mocks. We compute the mean signal-to-noise for our regular test sample and for the test sample rebuilt here. Figure~\ref{fig:snr} shows the histogram of these signal-to-noise ratios. We clearly see that the estimates from \cite{Perez-Rafols+2020a} have higher signal-to-noise, confirming our hypothesis.

To further test that changes to \squeze{} are responsible for an apparent decline in performance, we rerun the classification using the rebuilt samples to train and classify the code. For this run, the $f_1$ score including all objects down to magnitude $r=24.3$ is 0.91 for high-z quasars (with a confidence threshold of 0.30) and 0.47 for low-z quasars (with a confidence threshold of 0.29). This is significantly higher than our regular estimates and it is in agreement with the previous results from \cite{Perez-Rafols+2020a}. Thus, the discrepancy found here can be accounted for by the better signal-to-noise in the mocks used in the previous work (though we stress that the mocks we use here are more realistic).

\subsection{Explainability of the classifiers}
One of the main problems of using machine learning algorithms for classification problems is that often they are used as black boxes offering no explainability as to how the classification is being done. Because of the way \squeze{} is built, it provides some degree of explainability. What is more, the results of this paper also offer some degree of explainability to the classifiers presented in the previous papers in the series. Here we analyse \squeze{} training to review this.

To explain \squeze{} behaviour the most important thing is the coupling between classification and redshift estimation. The classification is done using a random forest classifier on a set of features. However, contrary to standard random forest usage, each spectrum can enter the random forest classifier multiple times. The key element here is that what we are classifying are not spectra, but trial redshifts. These are derived from the position of emission peaks found. This means that one of the key elements for \squeze{} to correctly identify the quasar spectra is its ability to detect real emission lines. Indeed, as shown in Figure~\ref{fig:f1_vs_mag}, our ability to detect the emission lines declines with increasing magnitude leading to worse performance.

Having confirmed that only emission peaks drive the classification (which is not necessarily the case in \citealt{Rodrigues+2023, Martinez-Solaeche+2023}), we now explore a feature importance analysis for each of the two random forest classifiers in \squeze{} in the four magnitude bins. This is performed by computing the mean (across the different trees in the forest) decrease in impurities when a particular feature is included or not. Higher values for the mean decrease indicate higher importance of the feature.

\begin{figure*}
    \centering
    \includegraphics[width=\textwidth]{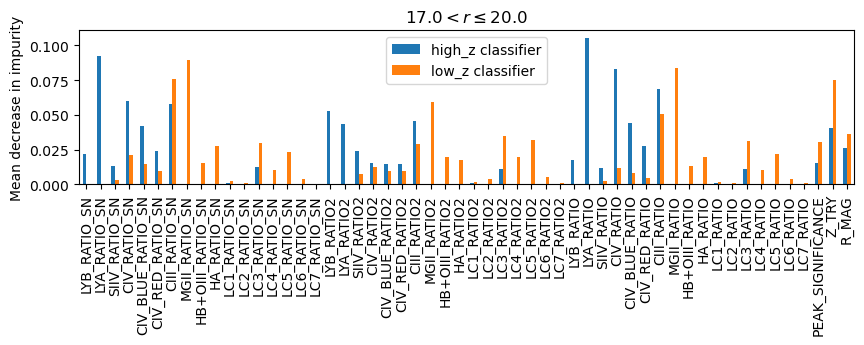}
    \includegraphics[width=\textwidth]{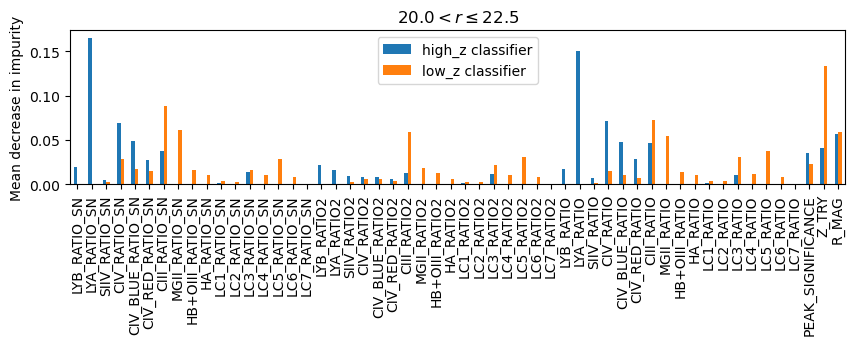}
    \includegraphics[width=\textwidth]{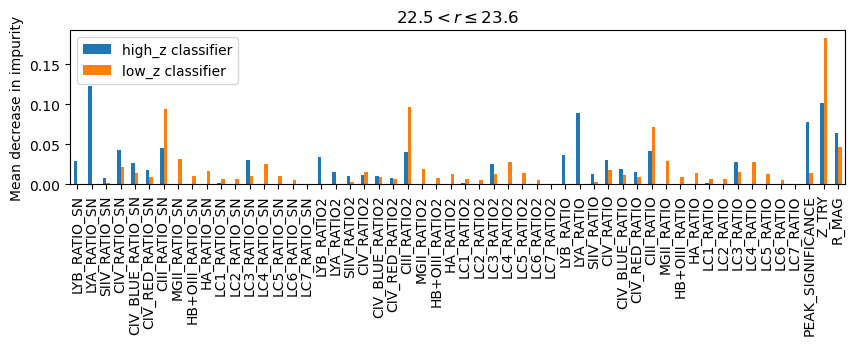}
    \caption{Feature importance analysis performed on \squeze{} training based on the mean decrease in impurity. Higher values indicate that the feature is more important. From top to bottom, the different panels show the result for the first three magnitude bins, with $17 < r \leq 20$, $20 < r \leq 22.5$, and t$22.5 < r \leq 23.6$. The results for the remaining magnitude bin are shown in figure~\ref{fig:feature_importance2}. }
    \label{fig:feature_importance}
\end{figure*}

\begin{figure*}
    \centering
    \includegraphics[width=\textwidth]{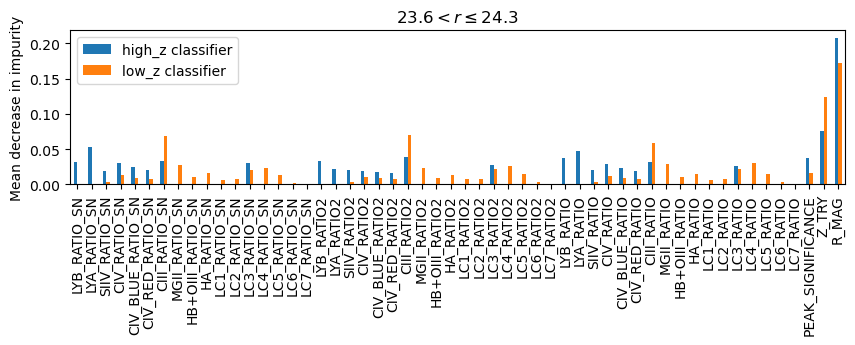}
    \caption{Same as Figure~\ref{fig:feature_importance} but for the fourth magnitude bin, with $23.6 < r \leq 24.3$.}
    \label{fig:feature_importance2}
\end{figure*}

The results of this exercise are shown in Figures~\ref{fig:feature_importance} and \ref{fig:feature_importance2}. Section~\ref{sec:squeze} describes 3 metrics for each of the lines (but see in more detail Equations 1 to 3 in \citealt{Perez-Rafols+2020b}). In terms of \squeze{} outputs, the amplitude of line X is labelled as X\_LINE\_RATIO, its significance as X\_LINE\_RATIO\_SN, and the slope at the base of the line as X\_LINE\_RATIO2. 

Generally speaking, the most important lines for high-z objects are \lya{}, \civ{} and \ciii{}, in this order. Their amplitudes are the most important characteristics, followed by the signal-to-noise. This makes sense as these are the most prominent emission lines, and these are the lines a human visual inspector usually looks for. This is suggestive that \squeze{} is indeed in agreement with the visual inspection analysis by a human expert. We also see that the red and blue halves of the \civ{} line are also relatively important. This line is the most affected by the presence of Broad Absorption Line (BAL) features and therefore these lines are important to include BAL quasars in our sample. 

As we go fainter, the importance of these emission lines decreases in favour of the trial redshift and the magnitude. Because we are no longer able to distinguish real emission line peaks from noise peaks, it seems that \squeze{} is relying more on the magnitude and redshift distribution of the objects. This explains why adding these two columns helps to improve the results (see Appendix~\ref{sec:add_cols}). The usage of these two columns is equivalent to having priors on their expected distributions. We warn the reader that this could potentially bias us towards the expected distributions and that we might be misled into overestimating the use of these parameters. 

Similarly, for low-z objects, the most important lines are \mgii{} and \ciii{}, in this order. Interestingly, the extra bands that we added to avoid line confusion, in particular, LC3, LC4 and LC5 (see Table~\ref{tab:line_confusion_intervals}) have a non-negligible importance. For instance, the mean decrease for these lines is similar to real lines such as the \ha{} or \hb{}+\oiii{}. They are also more important than the \lyb{}, \lya{}, \siiv{}, and \civ{} lines but this is expected as these are high-z lines.

This explains the improvement seen when changing the default set of lines (see Appendix~\ref{sec:new_lines}). As expected, when going fainter in magnitude we see a similar behaviour as for high-z candidates.

\subsection{Redshift precision}
In this Section, we discuss the estimated redshift precision of our catalogue and compare it with previous results from the literature. 

We already discussed the typical redshift errors of our correctly classified quasars in Section~\ref{sec:performance}, but there we analysed the entire sample, and not only the objects that would enter the quasar catalogue. Later, in Section~\ref{sec:quasar_cat}, we explained how we build our quasar catalogue. We follow the same procedure to build catalogues for the test and 1 deg$^{2}$ test samples, where the redshift and the classification are known, to estimate the precision of our redshift estimates. We quantify this precision in terms of $\sigma_{\rm NMAD}$ (see Equation~\ref{eq:NMAD}). 

We compute $\sigma_{\rm NMAD}$ considering only correct classifications for the entire sample, and for high-z quasars and low-z quasars separately. We also compute $\sigma_{\rm NMAD}$ for two bright sub-samples. For the first one, we consider $r<22.5$, i.e. including bins 1 and 2, where the emission lines are clearly detected (see Section~\ref{sec:performance}). For the second bright sample, we consider quasars with $r<21.3$, where we expect a much higher purity (in particular >0.90 for high-z quasars). 

Table~\ref{tab:z_precison} summarizes the results of this exercise. As expected, The results are similar in the test and 1 deg$^{2}$ test samples. In particular, the dispersion is lower for low-z quasars and also for brighter objects. Overall our results are below the per cent level. This is an order of magnitude better than the reported values of $\sigma_{\rm NMAD}=9\%$ by \cite{Matute+2012}, who analyse a similar sample of quasars in the ALHAMBRA survey down to magnitude $r=24$. These results are comparable to the findings by \cite{Chaves-Montero+2017}, where they get $\sigma_{\rm NMAD}=1.15\%$ and $\sigma_{\rm NMAD}=0.91\%$ for their AGN-X sample in the 2-line and 3-line detection mode, respectively. For their AGN-S sample, they find $\sigma_{\rm NMAD}=1.01\%$ and $\sigma_{\rm NMAD}=0.86\%$ for the 2-line and 3-line detection modes. However, we note that their quasars have magnitudes $F814W<22.5$ for the AGN-S sample and $F814W<23$ for the AGN-X sample. While there is no direct comparison between $r$-band magnitudes and F814W magnitudes, they have generally brighter quasars. If we compare their results to our brighter samples, we see that we recover $\sigma_{\rm NMAD}$ values that are $\sim5-20\%$ per cent lower (depending on the exact compared samples). 

\begin{table}
    \centering
    \caption{Normalised median absolute deviation, $\sigma_{\rm NMAD}$, of the correct classifications for the test and 1deg$^{2}$ test samples.}
    \label{tab:z_precison}
    \begin{tabular}{clcc}
\toprule
    sample & name &  $\sigma_{\rm NMAD}$ (per cent) &  $N$ \\
\midrule
\multirow{5}{*}{test}
& all &                    0.92 & 2694 \\
& high-z &                    0.99 & 1451 \\
& low-z &                    0.80 & 1243 \\
& $r<22.5$ &                    0.81 & 2075 \\
& $r<21.3$ &                    0.74 & 1289 \\\midrule
\multirow{5}{*}{1 deg$^2$ test}
& all &                    0.88 &  139 \\
& high-z &                    0.88 &   71 \\
& low-z &                    0.90 &   68 \\
& $r<22.5$ &                    0.79 &  102 \\
& $r<21.3$ &                    0.75 &   69 \\
\bottomrule
\end{tabular}
    \\
    \vspace{0.1cm}
    For each sample, we give the values for the entire sample, for high-z quasars and low-z quasars only, and for two bright samples with $r<22.5$ and $r<21.3$ respectively (see text for details).
\end{table}

\section{Summary and conclusions}\label{sec:summary}
In this paper, we analyzed miniJPAS data using \squeze{}. We presented the particularities of applying \squeze{} to this dataset and a catalogue of quasar candidates. Following previous papers in this series \citep{Rodrigues+2023, Martinez-Solaeche+2023}, we trained the models on the miniJPAS mocks developed by \cite{Queiroz+2022} for this purpose. We tested the performance on three different datasets, two of them synthetic and one of them using the relatively small subset of miniJPAS data with spectroscopic counterparts from SDSS. Finally, we compared our results to previous estimates of \squeze{} performance, attempted to explain the reasoning behind \squeze{}, assessed the impact of using different noise models to build the mocks, and evaluated the redshift precision of our samples. Our main conclusions are as follows:
\begin{itemize}
    \item Our results in the test samples suggest that the $f_1$ score including all objects down to $r= 24.3$ is 0.49 for high-z quasars and 0.24 for low-z quasars. For high-z quasars, this is increased to 0.9 for magnitudes $r < 21.0$.
    \item While \squeze{} performance is lower than some of the other classifiers of the series, it provides us with redshift estimates 
    \item We asses our redshift precision using the normalised median absolute deviation, $\sigma_{\rm NMAD}$. For our test sample, we reach a value of 0.92\%, an order of magnitude better than similar samples in the literature. For brighter samples, this decreases further to 0.81\% ($r<22.5$) and 0.74\% ($r<21.3$). 
    \item Contrary to other machine learning classifiers, the \squeze{} decisions can be followed: as we go fainter in magnitude \squeze{} is no longer able to distinguish real emission lines from noise peaks and more weight is given to the magnitude and redshift distributions.
    \item It is possible that \squeze{} is able to run on extended sources with similar levels of performance even if the training set only characterizes point-like sources. This could imply that the photometric properties of extended and point-like quasars are similar or that the criteria used to split between extended and point-like sources occasionally fails. 
    \item Changing the noise model used to create the mocks has an impact mostly at the faint end and sometimes results in lower redshift estimates.
    \item We computed a catalogue of quasar candidates for both point-like sources, with 301 candidates, and also including extended sources, with 1049 candidates. While extended sources are not included in our mocks, the comparison of the magnitude and redshift distributions of both catalogues suggests that \squeze{} could have a similar performance on extended objects compared to point-like objects.
    \item A spectroscopic follow-up of a large number of objects is crucial to verify the results of this work and could lead to improvements in the classifiers.
\end{itemize}

\section*{Acknowledgements}
This paper has gone through an internal review by the J-PAS collaboration. 

I.P.R. was supported by funding from the European Union's Horizon 2020 research and innovation programme under the Marie Sklodowskja-Curie grant agreement No. 754510. R.A.: acknowledges support from FAPESP, project 2022/03426-8. G.M.S. acknowledges financial support from the Severo Ochoa grant CEX2021-001131-S funded by MCIN/AEI/ 10.13039/501100011033., and to the AYA2016-77846- P and PID2019-109067-GB100. M.M.P. acknowledges support from the A*MIDEX project (ANR-11-IDEX-0001-02) funded by the ``Investissements d'Avenir'' French Government program, managed by the French National Research Agency (ANR), by ANR under contracts ANR-14-ACHN-0021 and ANR-22-CE31-0026 and by the Programme National Cosmology et Galaxies (PNCG) of CNRS/INSU with INP and IN2P3, co-funded by CEA and CNES. SB acknowledges support from the project PID2021-124243NB-C21 from the Spanish  Ministry of Economy and Competitiveness (MINECO/FEDER, UE) and partial support from the Project of Excellence Prometeo/2020/085 from the Conselleria d’Innovaci\'o, Universitats, Ci\`encia i Societat Digital de la Generalitat Valenciana. J.C.M. acknowledges financial support from the Spanish Ministry of Science, Innovation, and Universities through the project PGC2018-097585-B-C22 and the European Union’s Horizon Europe research and innovation programme (COSMO-LYA, grant agreement 101044612). VM thanks CNPq (Brazil) and FAPES (Brazil) for partial financial support. LSJ acknowledges the support from CNPq (308994/2021-3)  and FAPESP (2011/51680-6).

Based on observations made with the JST250 telescope and PathFinder camera for the miniJPAS project at the Observatorio Astrofísico de Javalambre (OAJ), in Teruel, owned, managed, and operated by the Centro de Estudios de Física del Cosmos de Aragón (CEFCA). We acknowledge the OAJ Data Processing and Archiving Unit (UPAD) for reducing and calibrating the OAJ data used in this work. Funding for OAJ, UPAD, and CEFCA has been provided by the Governments of Spain and Arag\'on through the Fondo de Inversiones de Teruel and their general budgets; the Aragonese Government through the Research Groups E96, E103, E16\_17R, E16\_20R and E16\_23R; the Spanish Ministry of Science and Innovation (MCIN/AEI/10.13039/501100011033 y FEDER, Una manera de hacer Europa) with grants PID2021-124918NB-C41, PID2021-124918NB-C42, PID2021-124918NA-C43, and PID2021-124918NB-C44; the Spanish Ministry of Science, Innovation and Universities (MCIU/AEI/FEDER, UE) with grant PGC2018-097585-B-C21; the Spanish Ministry of Economy and Competitiveness (MINECO) under AYA2015-66211-C2-1-P, AYA2015-66211-C2-2, AYA2012-30789, and ICTS-2009-14; and European FEDER funding (FCDD10-4E-867, FCDD13-4E-2685).



\bibliographystyle{aa}
\bibliography{main}


\begin{appendix}

\section{Effect of the noise model in mocks}\label{sec:noise}
As mentioned in Section~\ref{sec:data}, the mock sets used are described in detail in \cite{Queiroz+2022}. In particular, we used the noise model 11 which is precisely the one closest to the observed data. Here we explore the impact of using a different noise model. We train \squeze{} using the second-best noise model (model 1) to assess the performance on the respective test sample (computed also using the alternative noise model).

\begin{figure*}
    \centering
    \includegraphics[width=0.8\textwidth]{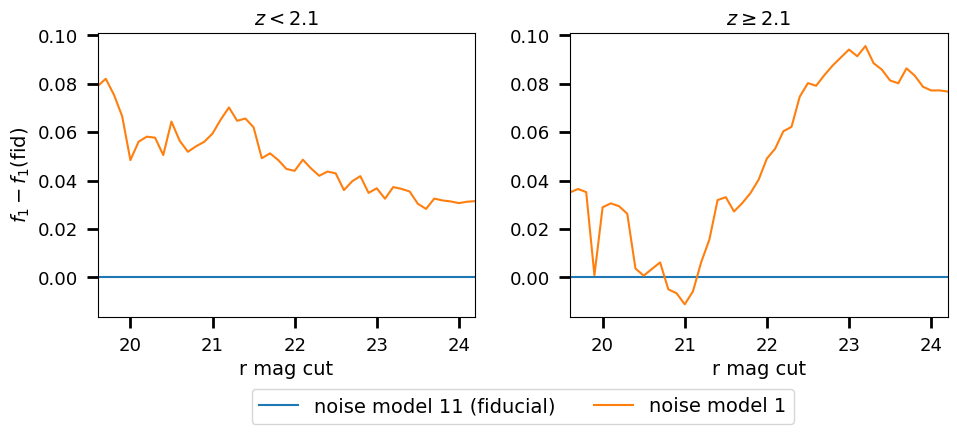}
    \caption{Change in the $f_{1}$ score when different noise models are used to generate the mocks. Note that the change is applied both to the train and the test samples. The left (right) panel shows the performance for the low (high) redshift quasars.}
    \label{fig:f1_noise_model}
\end{figure*}

Figure~\ref{fig:f1_noise_model} shows the change in $f_{1}$ score when the different noise models are used. The alternative noise model has slightly better performance. Including objects at all magnitudes, there is an increase in the performance of 0.08 for high-z quasars and 0.04 for low-z quasars. However, the noise model 1 is simpler than the 11th, and the test sample is also rerun for each noise model. Because noise model 1 is simpler, it is not unexpected that the performance is slightly better, as the classifiers have to learn a simpler distribution. However, we stress that noise model 11 is closer to the actual measurement of the noise distribution and thus should be more realistic (hence our choosing it as our fiducial model). 

\begin{figure*}
    \centering
    \includegraphics[width=0.8\textwidth]{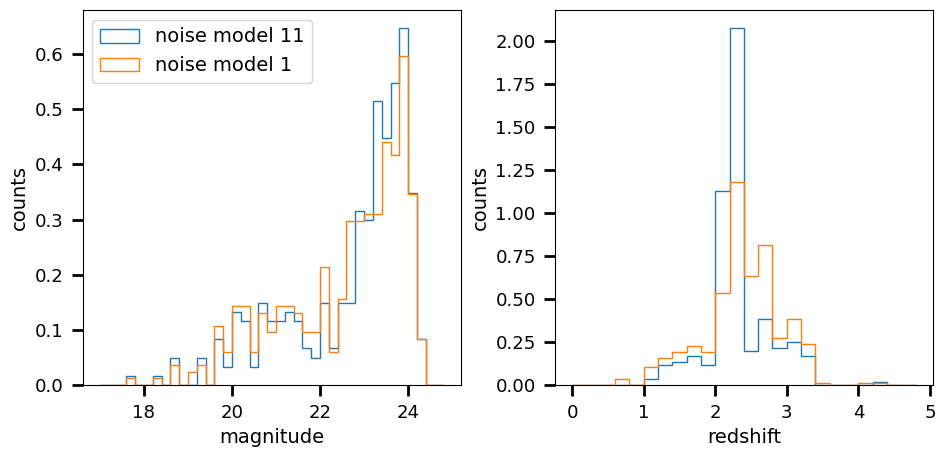}
    \caption{Distribution of magnitude (left) and redshift (right) for the quasar candidates computed using different noise models in the mocks to train our classifier. Noise model 11 is the closest to the observed data (see \protect\citealt{Queiroz+2022}).}
    \label{fig:data_dist_noise_model}
\end{figure*}

Perhaps more interesting is the impact of using these noise models on real data. Table~\ref{tab:data_dist_noise_model} shows the number of candidates recovered when using the different noise models to train the classifiers. Using model 11 results in a smaller number of candidates (by $\sim30\%$). A similar decrease is observed for both the point-like and the entire samples. We explore this difference further for the point-like sample by analyzing the distributions of redshift and magnitude (Figure~\ref{fig:data_dist_noise_model}). The magnitude distributions are essentially the same, but we observe some small differences in the redshift distribution. Model 1 seems to favour slightly larger redshifts. Clearly, one of the models has better redshift precision. One would tend to think that model 11, being a more realistic noise model, would produce better redshifts, but a spectroscopic follow-up of the objects is required to know for certain. 

\begin{table}
    \centering
    \caption{Number of quasar candidates in different noise models.}
    \label{tab:data_dist_noise_model}
\begin{tabular}{ccc}
\toprule
 noise model &  point-like candidates &  all candidates \\
\midrule
          11 &                    301 &                 1049 \\
           1 &                    419 &                 1514 \\
\bottomrule
\end{tabular}
    \\
    \vspace{0.1cm}
    Number of quasar candidates computed using different noise models in the mocks to train our classifier. Noise model 11 is the closest to the observed data (see \protect\citealt{Queiroz+2022}).
\end{table}

\section{Performance of the peak finders}\label{sec:peak_finder_comparison}

We stressed that one of the key elements of \squeze{} is the peak finder. In this work, we changed the original peak finder of \squeze{} by a new peak finder (see Section~\ref{sec:peak_finding}). Here we discuss the performance of the peak finders. 

\begin{figure}
    \centering
    \includegraphics[width=\columnwidth]{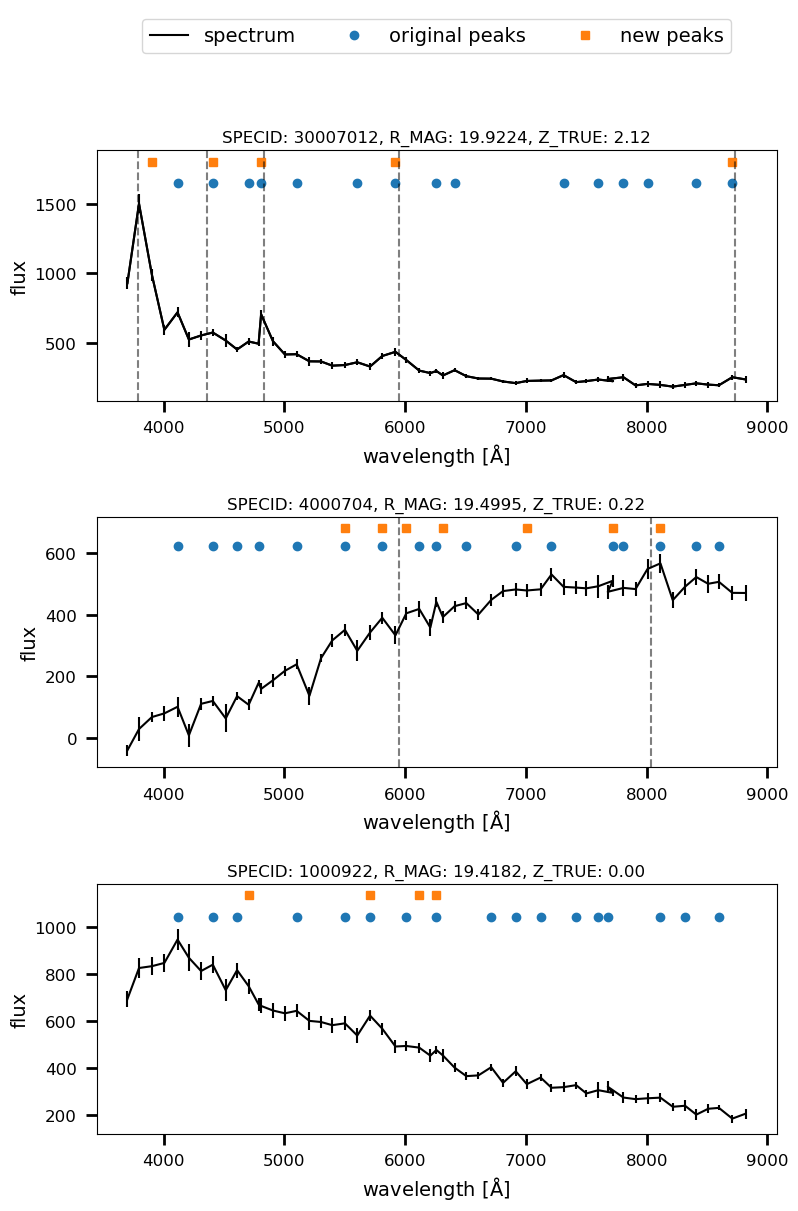}
    \caption{Example of the performance of the new peak finder compared to the original one. To illustrate the difference between peak finders, we show a quasar with magnitude $r=19.9224$ and redshift of $z=2.12$ (top panel), a galaxy with magnitude $r=19.4995$ and redshift of $z=0.07$ (mid panel) and a star with magnitude $r=19.4182$. Blue circles show the peaks as detected by the original peak finder. Orange squares are the peaks detected by the new peak finder. Dashed lines indicate the expected position of the main emission lines. For the quasar, from left to right, they indicate the \lya{}, \siiv{}, \civ{}, \ciii{} and \mgii{}. For the galaxy, also from left to right, they indicate the \oiii{} and \ha{}.}
    \label{fig:peak_finder}
\end{figure}

We start by performing a qualitative assessment based on the performance of a few bright objects (see Figure~\ref{fig:peak_finder}). The first example shows the performance of both the new peak finder and the original one on a synthetic spectrum of a quasar. We can see that the new peak finder provides a smaller number of peaks and that they are at the expected positions. The other two examples show the performance of synthetic spectra of a galaxy and of a star, where the assumption of a power-law continuum does not necessarily hold here. Nevertheless, the number of noise peaks is significantly reduced

These examples correspond to a relatively bright spectrum and the peak identification is less successful on fainter objects as the noise is larger. Nevertheless, we find that the new peak finder better filters the noise peaks. 

To estimate the performance of the peak finders we use three quantities. First, we consider the completeness after the peak finder step. As mentioned above, quasars for which we fail to detect the correct peak here will not be recovered at a later stage. Thus, we want this quantity to be as high as possible. Apart from completeness, it is also important to consider the number of correctly identified peaks and the total number of peaks. 

\begin{figure*}
    \centering
    \includegraphics[width=0.8\textwidth]{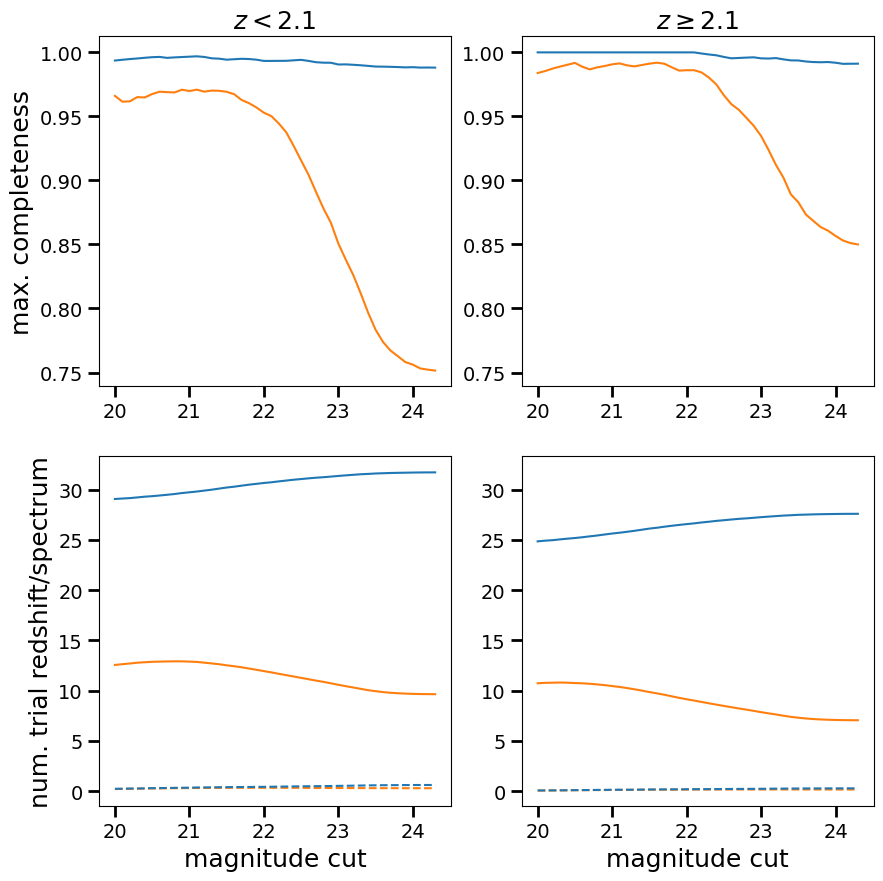}
    \caption{Top panels show the level of completeness after the peak finder step. The bottom panels show the number of peaks per spectrum as solid lines and the number of correct peaks per spectrum as dashed lines. To compute the later, only spectra of quasars are counted, whereas to compute the former all spectra are considered. The results for the old (new) peak finder are shown in blue (orange) lines.}
    \label{fig:peak_finder_performance_comp}
\end{figure*}

Figure~\ref{fig:peak_finder_performance_comp} shows the result of this exercise. The completeness of the new peak finder is lower than that of the original one. For high-z quasars, the decrease is 0.014 at magnitudes $r<22.1$, where the original peak finder has a completeness of 1. As we go fainter in magnitude, the difference increases up to 0.14 at magnitudes $r<24.3$, where the original peak finder completeness still remains 0.99. For low-z quasars, we see a similar trend albeit with a larger decrease: 0.043 at magnitudes $r<22.1$ and 0.23 at magnitudes $r<24.3$. 

This decrease in completeness is compensated by a drastic reduction in the number of incorrect peaks when the new peak finder is used. We see a decrease of a factor of $\sim3$ for high-z quasars and a factor of $\sim2$ for low-z quasars. At the same time, the number of correct peaks per spectrum stays roughly constant. This means that the random forest algorithms will have an easier job finding the correct entries. Indeed, we see an increase in performance when using the new peak finder (see Figure~\ref{fig:peak_finder}). We conclude that this decrease in the number of incorrect peaks more than compensates for the decrease in completeness. 

\begin{figure*}
    \centering
    \includegraphics[width=0.8\textwidth]{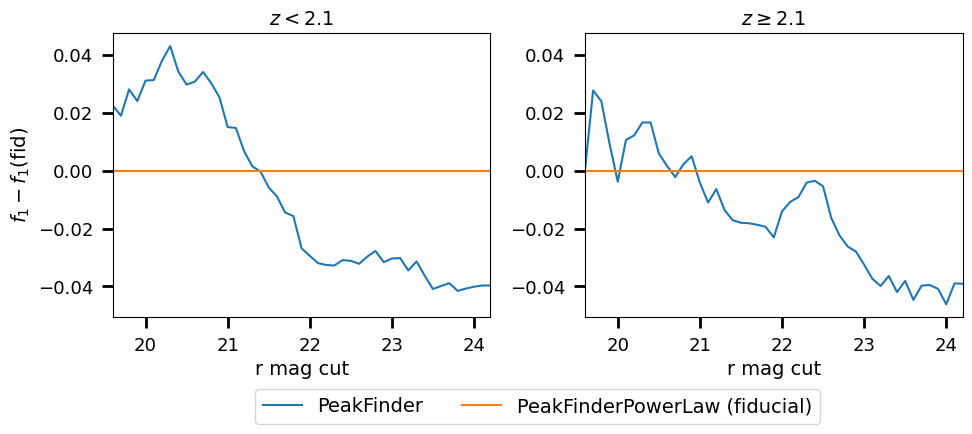}
    \caption{Changes in the performance ($f_1$ score) of SQUEzE when training with the different peak finders. The orange line shows the performance of the original peak finder, \texttt{PeakFinder}, compared to the peak finder developed here, \texttt{PeakFinderPowerLaw}, our fiducial choice. The left (right) panel shows the performance for the low (high) redshift quasars. For the original peak finder, we use the following parameters: 4 magnitude bins, no smoothing, a minimum significance of 0, using the default set of lines, using two random forests, and adding columns $z_{\rm try}$ and magnitude $r$. These parameters are selected following the approach described in Appendix~\ref{sec:robustness} for the new peak finder.} 
    \label{fig:peak_finder_comp}
\end{figure*}

\section{Robustness of the chosen \squeze{} parameters}\label{sec:robustness}
Section~\ref{sec:performance} shows the results of \squeze{} on the miniJPAS mocks. Then, in Section~\ref{sec:old} we discussed the reasons behind the decrease in the performance compared to the previous estimates from \cite{Perez-Rafols+2020a}. The performance decrease is attributed to the data being noisier than assumed by the rough estimates of \cite{Perez-Rafols+2020a}. In the subsequent Subsections, we review the main choices for the different parameters  and conclude that only minor tweaks to \squeze{} parameters are required in order to achieve the optimal configuration and that this configuration is only marginally better than the default choices. This supports the universality of \squeze{} models stated in \cite{Perez-Rafols+2020a}. 

Throughout this Section, we use the validation set to justify the different choices we make. We change one parameter at a time to evaluate the effect of changing this parameter, fixing the rest to our fiducial choice, as described in Section~\ref{sec:squeze}. Comparisons are made using the $f_1$ score as defined in Equation~\ref{eq:f1}. We checked that in all cases the purity and completeness remain roughly stable, i.e. that there is not a boost on the $f_1$ score by having a significantly higher purity at the cost of lower completeness or vice-versa. However, for clarity, here we only show the $f_1$ score.

\subsection{Training in magnitude bins}\label{sec:mag_bins} 
In Section~\ref{sec:classification} we classify objects separately based on their $r$-band magnitude. Here we explore the reasons behind this choice. The main argument to split the sample is that brighter objects have higher signal-to-noise and thus emission lines are easier to detect. On top of this, faint objects are substantially more numerous, thus dominating the training set. It is therefore reasonable to think that the random forest could learn to identify the low signal-to-noise quasars better and lower the performance at the bright end. 

To test if this is indeed the case, we run \squeze{} in three scenarios. First, we take all the objects in a single magnitude bin, $r\in\left(17.0, 24.3\right]$. Second, we split the bin in two, $r\in\left(17.0, 22.5\right]$ and $r\in\left(22.5, 24.3\right]$. Third, we further split each of the bins, $r\in\left(17.0, 20.0\right]$, $r\in\left(20.0, 22.5\right]$, $r\in\left(22.5, 23.6\right]$, and $r\in\left(23.6, 24.3\right]$. We train \squeze{} in each of these magnitude bins and combine the results into the larger single bin to compare. 

\begin{figure*}
    \centering
    \includegraphics[width=0.8\textwidth]{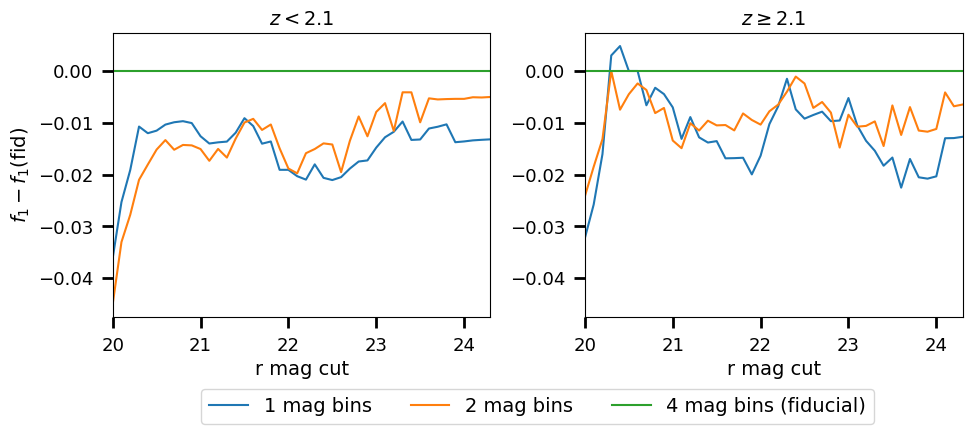}
    \caption{Changes in the performance ($f_1$ score) of SQUEzE when training with the different number of magnitude bins. The magnitude bins are $r\in\left(17.0, 24.3\right]$ for the model with 1 bin (blue lines), $r\in\left(17.0, 22.5\right]$ and $r\in\left(22.5, 24.3\right]$ for the model with 2 bins (orange lines), and $r\in\left(17.0, 20.0\right]$, $r\in\left(20.0, 22.5\right]$, $r\in\left(22.5, 23.6\right]$, and $r\in\left(23.6, 24.3\right]$ for the model with 4 bins (green lines). The left (right) panel shows the performance for the low (high) redshift quasars. The fiducial model, with 4 bins, is the most performant.}
    \label{fig:num_bins_comp}
\end{figure*}

Results of this exercise are shown in Figure~\ref{fig:num_bins_comp}, where we compare the performance of the models with 1 and 2 bins to that of the fiducial model, with 4 bins. When objects at all magnitudes are included, the performance drops for low-z quasars by $\sim0.015$ and $\sim0.005$ when using 1 or 2 bins, respectively. When cutting at different limiting magnitudes, we find the performance to be generally higher in the 4 bins model by $\sim0.02$ and $\sim0.01$ for low-z and high-z quasars, respectively. An even finer magnitude split might yield even better results, but larger amounts of data would be necessary, so we leave this for the future. 

\subsection{Significance threshold for peak finding}\label{sec:significance}
In Section~\ref{sec:peak_finding} we explained that peaks are identified by selecting outliers with a minimum significance from a power-law fit of the spectrum continua. Here we motivate the choice of $N=2$ in our standard runs. 

\begin{figure*}
    \centering
    \includegraphics[width=0.8\textwidth]{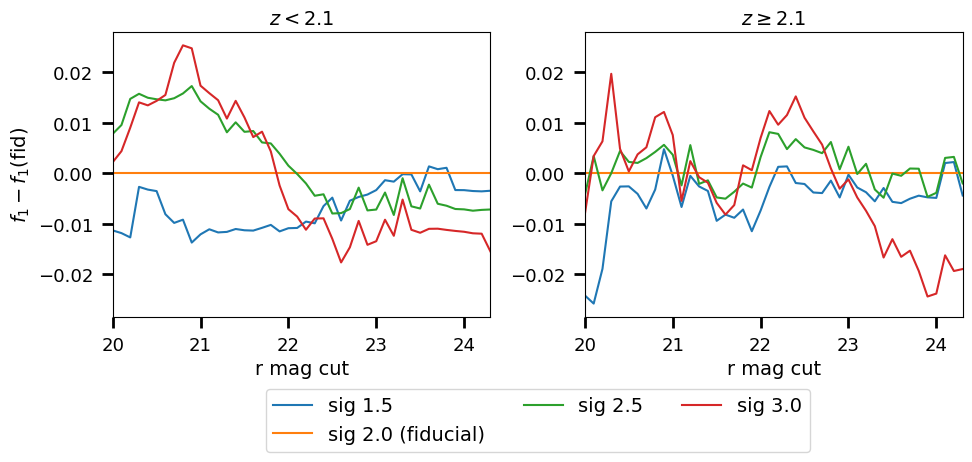}
    \caption{Comparison of the performance ($f_1$ score) of SQUEzE when the minimum peak significance is 1.5 (blue lines), 2.0 (orange lines), 2.5 (green lines), and 3.0 (red lines). The left (right) panel show the performance change for the low (high) redshift quasars. We see fluctuations in the performance at the percent level.}
    \label{fig:peak_sig_comp}
\end{figure*}

We compare the performance when taking different cuts in the peak significance. We explore a cut of 1.5 to 3.0 (both included) in steps of 0.5. We compare the performance of those models against the fiducial model, with a significance cut of 2.0. Results of this exercise are shown in  Figure~\ref{fig:peak_sig_comp}. The performance using different significant cuts is showing some fluctuations around a 0.01 change in performance. The fiducial choice seems to be marginally better than the other studied cases.

\subsection{Optimisation of the line bands}\label{sec:new_lines}
As shown in Section~\ref{sec:trial_z}, for each of the trial redshifts we compute a set of line metrics based on the predicted position of the emission lines of interest. The line bands used on the main results were optimized for BOSS data and our j-spectra have a significantly different resolution. Here we test if the same line bands should be used here, as seems to be indicated by the results in \cite{Perez-Rafols+2020a}. 

We change the line bands with the following criteria. First, we remove the weak emission lines that cannot be resolved in the mean spectrum of miniJPAS quasars (Martínez Uceta et al. In prep.). Then, we expand the bands to include more than one filter. This is important as there are non-detections in some of the filters, more so at the faint end. Having more than one filter in each of the bands allows us to measure the line metrics even when a few of these faulty measurements are present in the spectra. We label this set of lines as {\it wide}.

On top of widening the bands, we also add a few new ``emission lines''. These are added in intervals where we do not expect an emission line, but where we would expect it if we had line confusion (see Table~\ref{tab:line_confusion_new_lines}). With these, we not only increase the number of filters used to classify the j-spectra, but we also could potentially improve the existing line confusion. We label this set of lines as {\it wide+extra}.

\begin{figure*}
    \centering
    \includegraphics[width=0.35\textwidth]{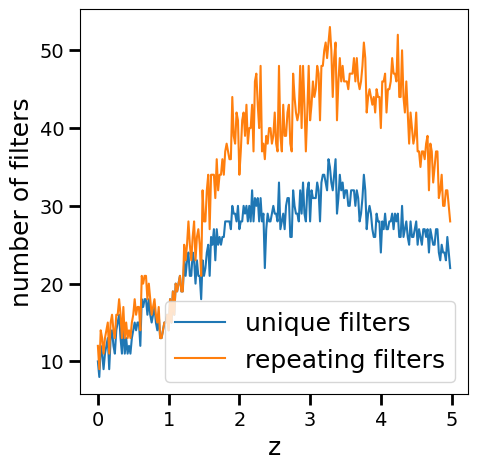}
    \hspace{0.05\textwidth}
    \includegraphics[width=0.35\textwidth]{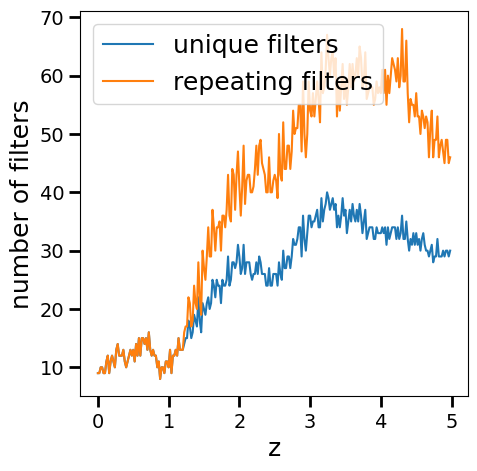}
    \caption{Same as Figure~\ref{fig:num_filters} but using the default bands (left panel) and using only the wider bands (right panel). In both cases use a smaller number of filters compared to the fiducial set of lines (Figure~\ref{fig:num_filters}).}
    \label{fig:new_lines_filters_used}
\end{figure*}

The chosen set of line bands is given in Table~\ref{tab:line_confusion_intervals}. To compare with the number of filters used in our fiducial choice (Figure~\ref{fig:num_filters}) we give this quantity for the default bands (left panel of Figure~\ref{fig:new_lines_filters_used}) and when we only use the wider bands (right panel of Figure~\ref{fig:new_lines_filters_used}). In both the {\it wide} and the {\it wide+extra} line bands the number of filters used is higher than in the default case (see Figure~\ref{fig:num_filters}).

\begin{table}
    \centering
    \caption{Intervals added to help with identified line confusions.}
    \label{tab:line_confusion_new_lines}
    \begin{tabular}{ccl}
    \toprule
    Real line & Assumed line & Interval(s) added \\
    \midrule
    \lya{} & \civ{} & LC1 (\mgii{}) \\     
    \civ{} & \ciii{} & LC1 (\mgii{}), LC5 (\ha{}) \\
    \civ{} & \lya{} & LC2 (\ciii{}), LC7 (\ha{})\\
    \ciii{} & \civ{} & LC2 (\ciii{}), LC6 (\ha{})\\
    \civ{} & \mgii{} & LC4 (\ha{})\\
    \mgii{} & \hb{} & LC4 (\ha{})\\
    \lya{} & \mgii{} & - \\
    \ciii{} & \hb{} & LC3 (\ha{})\\
    \bottomrule
    \end{tabular}
    \\
    \vspace{0.1cm}
    The first column gives the real line and the second column gives the assumed line. The third column gives the intervals added to help remove the line confusion. In parenthesis, we show the line whose peak is expected to appear in the interval if the assumed line was correct.
\end{table}

\begin{table*}
    \centering
    \caption{Alternative line bands used by \squeze{} to compute the metrics.}
    \label{tab:line_confusion_intervals}
    \begin{tabular}{cccccccc}
\toprule
     LINE &     WAVE &    START &      END &  BLUE\_START &  BLUE\_END &  RED\_START &  RED\_END \\
\midrule
      \lyb{} & 1,033.03 & 1,000.00 & 1,080.00 &      890.00 &    990.00 &   1,103.00 & 1,159.00 \\
      \lya{} & 1,215.67 & 1,190.00 & 1,260.00 &    1,103.00 &  1,159.00 &   1,295.00 & 1,350.00 \\
     \siiv{} & 1,396.76 & 1,370.00 & 1,417.00 &    1,295.00 &  1,350.00 &   1,432.00 & 1,494.50 \\
      \civ{} & 1,549.06 & 1,515.00 & 1,575.00 &    1,432.00 &  1,494.50 &   1,603.00 & 1,668.00 \\
 \civ{} blue & 1,549.06 & 1,515.00 & 1,549.06 &    1,432.00 &  1,494.50 &   1,603.00 & 1,668.00 \\
  \civ{} red & 1,549.06 & 1,549.06 & 1,575.00 &    1,432.00 &  1,494.50 &   1,603.00 & 1,668.00 \\
     \ciii{} & 1,908.73 & 1,865.00 & 1,940.00 &    1,756.00 &  1,835.00 &   1,970.00 & 2,060.00 \\
     \mgii{} & 2,798.75 & 2,720.00 & 2,816.00 &    2,570.00 &  2,690.00 &   2,851.00 & 2,984.00 \\
  \hb{}+\oiii{} & 4,862.68 & 4,800.00 & 5,020.00 &    4,640.00 &  4,740.00 &   5,060.00 & 5,145.00 \\
       \ha{} & 6,564.61 & 6,480.00 & 6,650.00 &    6,320.00 &  6,460.00 &   6,750.00 & 6,850.00 \\
\midrule
      LC1 & 2,233.88 & 2,140.00 & 2,315.00 &    2,060.00 &  2,110.00 &   2,335.00 & 2,385.00 \\
      LC2 & 2,392.05 & 2,280.00 & 2,470.00 &    2,200.00 &  2,260.00 &   2,490.00 & 2,550.00 \\
      LC3 & 2,576.78 & 2,520.00 & 2,630.00 &    2,420.00 &  2,500.00 &   2,640.00 & 2,710.00 \\
      LC4 & 3,705.86 & 3,600.00 & 3,790.00 &    3,510.00 &  3,580.00 &   3,800.00 & 3,900.00 \\
      LC5 & 5,327.61 & 5,280.00 & 5,400.00 &    5,160.00 &  5,250.00 &   5,480.00 & 5,570.00 \\
      LC6 & 8,088.82 & 7,980.00 & 8,230.00 &    7,870.00 &  7,960.00 &   8,280.00 & 8,550.00 \\
      LC7 & 8,364.91 & 8,280.00 & 8,550.00 &    8,120.00 &  8,260.00 &   8,650.00 & 8,790.00 \\
\bottomrule
\end{tabular}
    \\
    \vspace{0.1cm}
    In the first block, intervals of major lines have been widened to ensure continuous coverage at a given range of redshift. Minor lines (e.g. NeIV and Ne V) have been removed. OIII and H$\beta$ lines have been merged into a single wider line. In the second block, additional line intervals have been added to help deal with identified line confusion (see Table~\ref{tab:line_confusion_new_lines}). All quantities are given in \angs{}.
\end{table*}

\begin{figure*}
    \centering
    \includegraphics[width=0.8\textwidth]{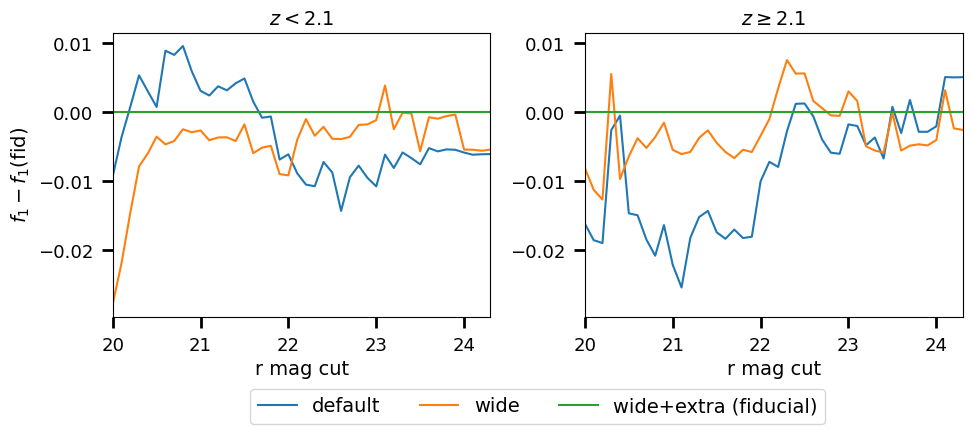}
    \caption{Comparison of the performance ($f_1$ score) of \squeze{} when using the default set of line bands and the two new sets of line bands (using wider bands and using both wider bands and some extra bands), specified in Table~\ref{tab:line_confusion_intervals}. Both cases are computed with a minimum peak significance of 2. The left (right) panel shows the performance for the low (high) redshift quasars. The fiducial model, using the wide+extra line bands is the most performant.}
    \label{fig:new_lines_comp}
\end{figure*}

We test the performance of the new sets of lines (using only the wider bands and using both the wider and extra bands) and compare it to the default lines. This comparison is shown in Figure~\ref{fig:new_lines_comp}. The {\it wide+extra} lines are superior, but only marginally, with an increase of order 0.01-0.02, seen mostly for bright magnitudes. This fact supports the predictions by \cite{Perez-Rafols+2020a} on the universality of their model, where only marginal improvements are expected when changing the parameters of the model.

\subsection{Single classifier vs redshift split classifier}\label{sec:new_rf}
\begin{figure*}
    \centering
    \includegraphics[width=0.8\textwidth]{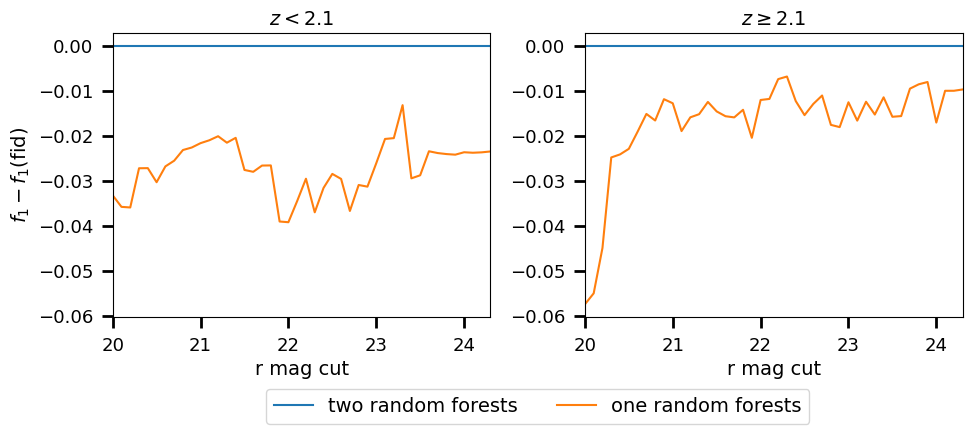}
    \caption{Comparison of the performance ($f_1$ score) of \squeze{} when using one or two random forest classifiers. When two are used, the first one is charged with classifying the low-z quasars and the second one, the high-z quasars. 
    The left (right) panel shows the performance for the low (high) redshift quasars. The fiducial model, using two random forests, is the most performant.}
    \label{fig:new_rf_comp}
\end{figure*}

In \cite{Perez-Rafols+2020b}, they argue that using two different random forest classifiers, one for high redshift $z_{\rm try}$ and one for low redshift $z_{\rm try}$, they obtain a better performance as opposed to using a single classifier. Since our datasets have significantly different properties (in particular magnitude and resolution), this statement does not necessarily hold here. In order to check this, we redo our analysis using a single random forest classifier. The options passed to \squeze{} are \texttt{\{"criterion": "entropy", "max\_depth": 10, "n\_jobs": 3, "n\_estimators": 1000\}}. We compare the results from this new run with our fiducial in Figure~\ref{fig:new_rf_comp}. There is a consistent increase of $\sim0.03-0.04$ in the performance at low redshift when using two random forest classifiers. For high redshift quasars, the improvement in the performance is smaller, reaching only $\sim0.02$. Thus, we decided to stick with using two random forest classifiers.

\subsection{Impact of redshit tolerance}\label{sec:z_tolerance}
Section~\ref{sec:performance} shows that as we go fainter in magnitude, the redshift precision decreases. This suggests that the redshift tolerance used could be too large. On the other hand, \citep{Perez-Rafols+2020a} used an even larger redshift tolerance (0.15). Here we discuss the possible effect of using different cuts on the redshift tolerance. We run \squeze{} using $\Delta z = \left|z_{\rm true} - z_{\rm try}\right| < 0.15$,  $\Delta z < 0.10$ (our fiducial choice) and $\Delta z < 0.05$.  

\begin{figure*}
    \centering
    \includegraphics[width=0.8\textwidth]{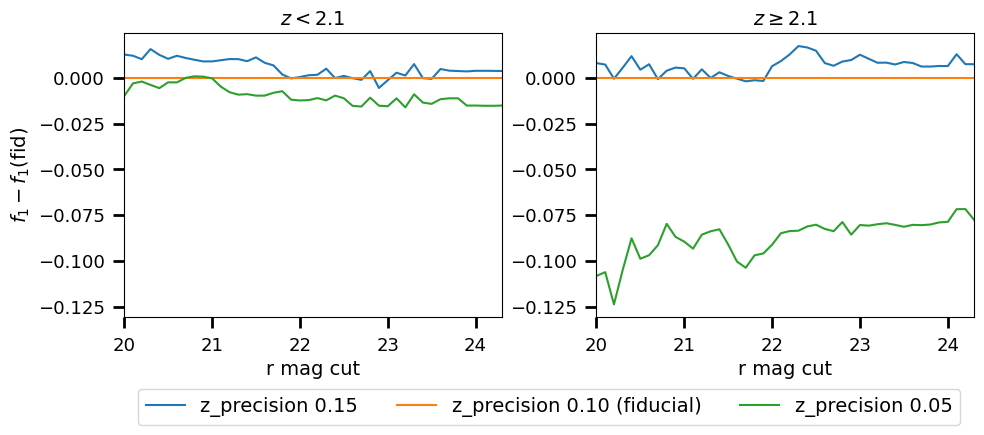}
    \caption{Comparison of the performance ($f_1$ score) of \squeze{} when using different redshift requirements to define correct classifications. Our fiducial choice is 0.10. The left (right) panel shows the performance for the low (high) redshift quasars. The performance of the fiducial model, with $\Delta z=0.10$, is marginally worse than the performance with $\Delta z=0.15$, but the redshift errors are significantly smaller (see Table~\ref{tab:z_dist_dz}).}
    \label{fig:z_precision}
\end{figure*}

Table~\ref{tab:z_dist_dz} shows the mean redshift offset and the dispersion measured in each of these three cases. As expected, the redshift precision increases as tighter constraints in $\Delta z$ are used. However, when the constraints are too tight, we start to impact the performance. 
For instance, in Figure~\ref{fig:z_precision} shows that using $\Delta z$ of 0.10 and 0.15 results in very similar performance levels, but that using $\Delta z $ of 0.05 implies a performance drop of $\sim0.10$ for the high-z quasars. Naturally, the comparison here is not so simple, as the criteria for correct classifications change with the chosen $\Delta z$. While this is true, this should only affect those quasars in which the trial redshift is close to the true redshift. In cases where line confusion occurs, then the chosen $\Delta z$ is irrelevant as the redshift error is much larger. The line confusion plots (Figure~\ref{fig:line_confusion}) indicate that the latter is driving the performance. We conclude that the fiducial $\Delta z=0.10$ is the optimal threshold choice as it has a similar performance as choosing $\Delta z=0.15$, but with better redshift precision.

\begin{table*}
    \centering
    \caption{Statistics of the redshift precision for different redshift precision requirements.}
    \label{tab:z_dist_dz}
    \begin{tabular}{cccccc}
\toprule
$\Delta z$    &     mag bin &  $\overline{\Delta v}$ (km/s) &  $\sigma_{\Delta v}$ (km/s) &  $\sigma_{\rm NMAD}$ (per cent) &  $N$ \\
\midrule
\multirow{4}{*}{0.15}
& $17.0 < r \leq 20.0$ &                        224.30 &                    4,282.73 &                    0.77 &  520 \\
& $20.0 < r \leq 22.5$ &                        968.87 &                    3,834.35 &                    0.99 & 2209 \\
& $22.5 < r \leq 23.6$ &                      1,349.70 &                    6,016.32 &                    1.66 & 1061 \\
& $23.6 < r \leq 24.3$ &                      1,603.68 &                    6,355.87 &                    1.86 &  210 \\
\midrule
\multirow{4}{*}{0.10}
& $17.0 < r \leq 20.0$ &                        457.03 &                    2,870.53 &                    0.74 &  517 \\
& $20.0 < r \leq 22.5$ &                        711.28 &                    3,121.52 &                    0.95 & 2198 \\
& $22.5 < r \leq 23.6$ &                      1,015.71 &                    4,209.44 &                    1.42 &  957 \\
& $23.6 < r \leq 24.3$ &                        889.00 &                    4,802.76 &                    1.55 &  184 \\
\midrule
\multirow{4}{*}{0.05}
& $17.0 < r \leq 20.0$ &                        463.10 &                    2,188.32 &                    0.69 &  491 \\
& $20.0 < r \leq 22.5$ &                        461.34 &                    2,274.05 &                    0.79 & 1998 \\
& $22.5 < r \leq 23.6$ &                        352.09 &                    2,642.76 &                    1.12 &  747 \\
& $23.6 < r \leq 24.3$ &                          5.51 &                    2,624.00 &                    1.10 &  125 \\
\bottomrule
\end{tabular}
    \\
    \vspace{0.1cm}
    Mean redshift offset and dispersion for the different magnitude bins. The mean offset indicates potential biases of our redshift estimate and the dispersion indicates our typical redshift error. Each block shows the results when running with different values for $\Delta z$. We note that our fiducial model has a significance $\Delta z = 0.15$.
\end{table*}

\subsection{Passing extra features to the random forest classifiers}\label{sec:add_cols}
\begin{figure*}
    \centering
    \includegraphics[width=0.8\textwidth]{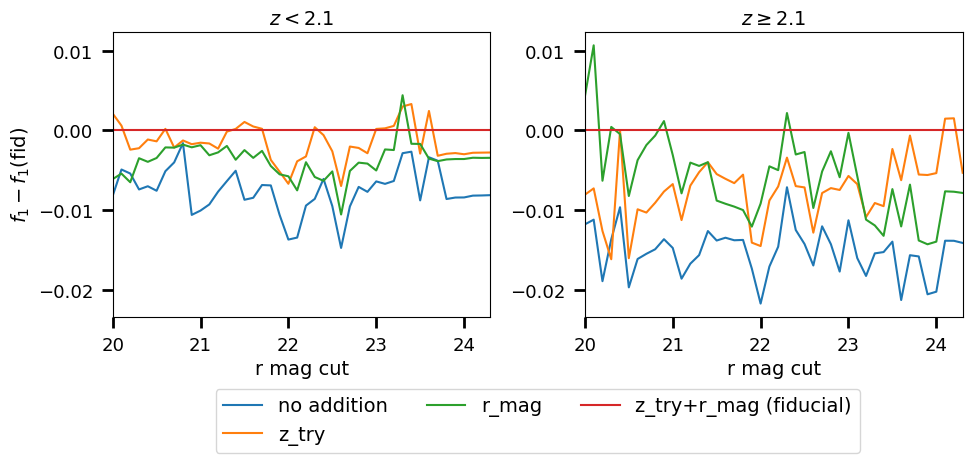}
    \caption{Comparison of the performance ($f_1$ score) of \squeze{} when we feed different columns to the random forests. We compare \squeze{} default choice, i.e. only feeding the metrics with the cases where we also feed it the trial redshift, the r-band magnitude, or both (our fiducial choice). The left (right) panel shows the performance for the low (high) redshift quasars. The fiducial model is the most performant. }
    \label{fig:add_cols}
\end{figure*}

To finalize the revision of our choices, we assess the impact of adding extra features to the random forest classifiers. In particular, we explore adding the trial redshift and/or the r-band magnitude to the list of parameters fed to the classifiers. We compare our fiducial choice (adding both parameters) with the cases where only one of the two features is passed, and when none is. The results of this exercise, shown in Figure~\ref{fig:add_cols}, indicate that the code is indeed performing optimally when both features are added. However, the performance change is at the per cent level, as in previous cases.

\end{appendix}


\end{document}